\documentclass[aps,prd,twocolumn,showkeys,superscriptaddress,groupedaddress,nofootinbib]{revtex4-1}  
\pdfoutput=1
\usepackage{graphicx}  
\usepackage{dcolumn}   
\usepackage{bm}        
\usepackage{amssymb}   
\usepackage{amsmath}
\usepackage{amsfonts}
\usepackage{slashed}
\usepackage{color}
\usepackage{mathtools} 
\usepackage{hyperref}
\usepackage{slashed, latexsym, verbatim, cancel}
\usepackage{pstricks}
\usepackage{bbm}
\usepackage{enumitem} 
\begin{document}


\title{Identifying a Higgsino-like NLSP in the context of a keV-scale gravitino LSP}

\author{Juhi Dutta}
\email{juhidutta@hri.res.in}
\affiliation{Regional Centre for Accelerator-based Particle Physics, Harish-Chandra Research Institute, HBNI, Chhatnag Road, Jhusi, Allahabad 211019, India}
\author{Biswarup Mukhopadhyaya}
\email{biswarup@hri.res.in}
\affiliation{Regional Centre for Accelerator-based Particle Physics, Harish-Chandra Research Institute, HBNI, Chhatnag Road, Jhusi, Allahabad 211019, India}
\author{Santosh Kumar Rai}
\email{skrai@hri.res.in}
\affiliation{Regional Centre for Accelerator-based Particle Physics, Harish-Chandra Research Institute, HBNI, Chhatnag Road, Jhusi, Allahabad 211019, India}

\date{\today}
\begin{abstract}
The presence of a Higgsino-like neutralino next-to-lightest sparticle (NLSP) and a keV scale gravitino ($\widetilde{G}$) LSP opens 
up new decay modes of the NLSP,  mainly to a Higgs/$Z$ boson and the LSP. Besides, a keV-scale gravitino as a 
warm dark matter candidate salvages a relatively light Higgsino-like NLSP from dark matter constraints. We focus on the prospects of observing 
 at least one b-jet and two opposite sign leptons with large missing transverse momenta ($\geq 1 b$ $+$ $ \ell^+\ell^-+\slashed{E}_T$)
 signal at the LHC.  
A distinguishing feature of this scenario is the production of longitudinal $Z$ bosons in neutralino decays, unlike in the case of gauginolike 
neutralinos, where the 
$Z$ is mostly transverse. The polarization information of the parent $Z$ boson gets reflected
in the angular distributions of the decay leptons and in some other variables derived therefrom.
\end{abstract}

\maketitle

\section{\label{sec:level1}Introduction}
In the light of the 13 TeV LHC run 2 results \cite{ATLAS,CMS}, both experimentalists and theorists are
leaving no stone unturned to interpret results of different supersymmetric (SUSY) scenarios, primarily to ensure no
glaring gaps are left in the current search strategies such that the signals may slip through them. 
Although bounds on sparticle masses from LHC are steadily increasing, they are derived in the context of simplified scenarios.
However, generic features on collider signals are helpful to investigate, if such features reflect the spectrum and the 
composition of the superparticle states.
 
The $R$-parity conserving minimal supersymmetric standard model (MSSM) ensures a stable, neutral, colorless dark matter (DM) candidate, 
generally the lightest neutralino ($\widetilde{\chi}^0_1$) which is the lightest SUSY particle (LSP). Depending on its 
composition, it could be either a democratic admixture of gaugino and Higgsino states or dominantly binolike, winolike 
or Higgsino-like. The DM composition faces stringent constraints from direct search data and indirect evidence as well as from 
relic density measurements by Planck \cite{Aghanim:2018eyx}. For example, a bino-dominated $\widetilde{\chi}^0_1$ 
with an appropriate small admixture of Higgsinos can by and large be consistent with relic density as well as 
direct search constraints. However for a winolike LSP, SU(2)-driven annihilation channels lead to strong constraints from $\gamma$-ray \cite{Fan:2013faa}
as well as positron data \cite{Abdughani2019}. 
A light Higgsino-like LSP is disfavored from direct dark matter searches \cite{Arcadi2018,Baer:2018rhs}.
 Conversely, the relic density being inversely proportional to the annihilation cross section, 
shows an underabundance for wino or Higgsino-like LSP whereas 
a binolike LSP can lead to an overabundance unless coannihilation occurs at an adequate rate. 
Mixed bino-Higgsino or bino-wino dark matter scenarios could 
 be more consistent from these standpoints \cite{Abdughani:2017dqs}.
The presence of a lighter particle such as a gravitino or an axino as the lightest SUSY particle relaxes these constraints on the
composition of the lightest neutralino $\widetilde{\chi}^0_1$, 
which now serves as the next-to-lightest sparticle (NLSP). Gravitinos are the LSP in models like gauge mediated supersymmetry breaking with
low scale of SUSY breaking. Here, the gravitino mass scale may be lighter than the MSSM sparticle masses if the scale of SUSY breaking is light enough to allow a light
gravitino mass governed by \cite{Martin:1997ns,Drees:2004jm,Giudice:1998bp},

\begin{equation}
m_{\widetilde{G}} = \frac{<F>}{\sqrt{3}M_{Pl}}
\end{equation}
where $m_{\widetilde{G}}$ is the mass of the gravitino, $<F>$ is the SUSY breaking scale 
and $M_{Pl}$ is the Planck scale.
This means that depending on the SUSY breaking scale $<F>$, the gravitino can be as 
light as $\mathcal{O}$(keV). A light gravitino with mass of few keV is motivated to be a warm 
dark matter candidate \cite{Baur:2015jsy,Boyarsky_2009,LCovi}. In this work we consider 
the LHC signals of a Higgsino-dominated NLSP where one has such a light gravitino LSP. 
 
Signals for a dominantly Higgsino-like $\widetilde{\chi}^0_{1}$ NLSP with a gravitino LSP 
has been studied by experimental collaborations at the LHC. A primarily Higgsino-like 
$\widetilde{\chi}^0_{1}$ NLSP is found to decay dominantly to either the Higgs or $Z$ 
boson along with gravitino. Since the light standard model (SM) like Higgs has the largest 
decay probability to $b \, \bar{b}$,
this leads to a final state dominated by hard $b$-tagged jets along with 
large missing transverse momentum (MET) $\slashed{E}_T$ and additional light 
jets/leptons arising from an accompanying $Z$ boson. 
Signatures for the Higgsino-like NLSP's have been studied in the context 
of Tevatron \cite{Meade:2009qv,Matchev:1999ft} and also at LHC where both 
CMS \cite{Sirunyan:2018ubx} and ATLAS \cite{ATLASGGM} have looked at multiple 
$b$ jets + MET, dilepton, and multilepton states to constrain a Higgsino-like NLSP scenario. 

In this work, we aim to study signatures of a low-lying Higgsino sector in the presence 
of a light gravitino LSP with emphasis on determining how the NLSP nature can be 
convincingly identified. To do this one would like to reconstruct the decay products of the
NLSP. As the Higgsino NLSP would decay to a light Higgs or a $Z$ boson, we may 
be able to observe their properties by appropriately reconstructing the Higgs boson through the 
$b$ jets arising from its decay as well as the $Z$ boson through the opposite sign 
dilepton pair from the gauge boson's decay respectively. We note here a very important 
and interesting feature of the decay of the NLSP.  It is expected that the $Z$ boson 
arising from the Higgsino-like $\widetilde{\chi}^{0}_1$ decay would be 
dominantly \textit{longitudinal} (Goldstone boson), primarily following the 
{\it equivalence theorem} where, after electroweak symmetry breaking the neutral 
Goldstone boson constitutes the longitudinal mode of the $Z$ boson responsible for its 
mass. This property if observed in the decay of the NLSP would exclusively point 
towards the presence of a Higgsino-like $\widetilde{\chi}^{0}_1$, helping us identify 
the nature of the NLSP. The direct production of the electroweak neutralino NLSP 
would be limited at the LHC as their mass becomes larger. However, the property of the
NLSP could still be studied if they are produced in cascade decays of strongly interacting 
sparticles. We therefore study the effect of including the strong sector in exploring 
the compositions of the NLSP as well as from the direct production of the 
low-lying electroweakinos (still allowed by experiments) and propose some new 
kinematic observables which help identify the NLSP. Thus, the salient points of our study 
are as follows:
\begin{enumerate}[label=(\roman*)]
 \item We consider a naturally compressed low-lying Higgsino sector as well as partially 
 and$/$or fully compressed spectra with the strongly interacting sparticles sitting above the NLSP. 
 The sparticles decay via cascades to the NLSP which further decays to a Higgs and a
  $Z$ boson thereby giving rise to  at least 1 $b$ jet and opposite-sign same flavor dileptons along with 
  missing transverse energy in the final state.
 
\item The characteristic features of a longitudinal $Z$ boson arising from decay of the Higgsino-like 
$\widetilde{\chi}^0_1$ are studied by utilizing angular variables of the negatively charged lepton.  In 
order to distinguish it from transversely polarized $Z$ bosons coming from other sources, we compare our 
results with the complementary admixture of NLSP, especially gaugino-dominated neutralinos as well as 
the SM background. 
 
 \item We observe that for a spectrum with a heavy NLSP, reflecting overall compression with respect to the strong 
 sector leads to an increased fraction of the longitudinal mode in the $Z$ boson arising from the NLSP decay.
 \item New observables enhancing the asymmetry in the angular distributions of the negatively charged lepton 
 have been proposed in order to characterize a longitudinally polarized $Z$ boson in comparison to a 
 transversely polarized $Z$ boson. Such observables distinctly vary depending 
 on the Higgsino-gaugino admixture of the NLSP and crucially capture the effect of the equivalence theorem for a heavy NLSP. 
 \end{enumerate}

The paper is organized as follows. In Sec. II, we discuss the current scenario with the Higgsino NLSP and gravitino LSP followed by the decay properties of the Higgsinos in Sec. III. In Sec. IV, we discuss the experimental status of a Higgsino-like NLSP with a light gravitino LSP at LHC. In Sec V, we choose some benchmarks to study the available parameter space. We perform the collider study and discuss our results at the high luminosity run of LHC in Sec. VI. In Sec. VII, we distinguish between the features of longitudinal and transverse gauge bosons. Sec. VIII summarizes the main conclusions of our work.

\section{Higgsino-dominated NLSP with keV LSP}

In our work, we discuss light Higgsino-like NLSP as a possible consequence of a general phenomenological MSSM. Since no hint of SUSY has yet shown up at direct searches, various possible configurations of the lightest neutralino, $\widetilde{\chi}^0_1$,
leading to distinct signals at colliders are of interest. Such a light Higgsino-like $\widetilde{\chi}^0_1$ is characterized by a low ($\lesssim 800$ GeV) $\mu$ parameter and heavy bino, wino soft mass parameters, i.e, $|\mu|<<M_1,M_2$. $\mu$ in the aforesaid range is also a preferred choice from the angle of naturalness \cite{Baer:2012cf,Baer:2013ava,Mustafayev:2014lqa,Baer:2012up,Baer:2016usl,Baer2017}. However $\widetilde{\chi}^0_1$, may not be the LSP in many situations. In such cases, there can be several other candidates for LSP such as gravitinos, axinos, sneutrinos, etc. 
 
The gravitino is the spin $\frac{3}{2}$ superpartner of the spin 2 graviton in local SUSY. Upon spontaneous SUSY breaking, there arises a massless Weyl fermion known as the goldstino
($\widetilde{G}$), owing to the breaking of the fermionic generators of
SUSY. After electroweak symmetry breaking, the gravitino acquires mass by absorbing the goldstino, which form the spin $\frac{1}{2}$ components of 
the massive gravitino. We henceforth approximate a light gravitino by the goldstino using the equivalence theorem.

The goldstino ($\widetilde{G}$) Lagrangian is~\cite{Martin:1997ns},
\begin{equation}
 \mathcal{L}_{goldstino} = i\widetilde{G}^\dagger  \bar{\sigma}^{\mu}  \partial_{\mu}\widetilde{G}  - \frac{1}{<F>}\widetilde{G}\partial_{\mu}j^{\mu}+c.c., 
\end{equation}
where $j^{\mu}$ 
 refers to the supercurrent involving all other sparticles and SM particles. 
The couplings of the gravitino to fermion-sfermion, gauge boson-gauginos are computed in Ref.~\cite{Baer:2006rs}.  
Having said this, we briefly discuss the couplings and decays of a Higgsino-like NLSP in the presence
of a light $\widetilde{G}$ before moving on to our numerical analysis.

\section{Higgsino NLSP decays}
\label{Decays}
A Higgsino-like $\widetilde{\chi}^0_1$ is characterized by a large Higgsino fraction with suppressed wino  and bino fractions, i.e, $\mu< M_1, 
M_2$. In the presence of a light $\widetilde{G}$ LSP,  the Higgsino-like $\widetilde{\chi}^0_1$ NLSP decays to either a Higgs ($h$) 
or a $Z$ boson and $\widetilde{G}$. 
Absence of a large bino component leads to a rather 
suppressed photon mode unless there is substantial gaugino-bino-Higgsino admixture \cite{Matchev:1999ft}. However the photon mode may dominate
in case of very light Higgsinos, where the decay to the Higgs or $Z$ boson is phase space suppressed. 
As the coupling of a gravitino to other particles are inversely proportional to its mass($m_{\widetilde{G}}$), a lighter gravitino has stronger couplings as compared to a heavier one.
  For any sparticle $\widetilde{X}$ decaying into its SM partner $X$ and the gravitino, the width is 
  given by \cite{Martin:1997ns}: 
\begin{equation}
\label{eq1}
  \Gamma (\widetilde{X} \rightarrow X \widetilde{G}) =
  \frac{m^5_{\widetilde{X}}}{48 \pi M^2_P m^2_{\widetilde{G}}} ( 1-\frac{m^2_{X}}{m^2_{\widetilde{X}}})^4
\end{equation}
where $m_{X(\widetilde{X})}$ refers to the mass of $X(\widetilde{X})$. As we are interested in the decay of the 
neutralino NLSP to the gravitino, the composition of the lightest neutralino becomes an essential characteristic as 
it would determine what the NLSP finally decays to. 
The neutralino mass matrix in the basis ($\widetilde{B}, \widetilde{W_3}, \widetilde{H}^0_d,\widetilde{H}^0_u$) is as follows \cite{Martin:1997ns}:
\begin{equation}\scriptsize
M^{n} =  \left( \begin{array}{cccc}
M_{1} & 0 & -M_{Z}s_W c_\beta & M_{Z}s_W s_\beta\\
0 &  M_{2} & M_{Z}c_W c_\beta & -M_{Z}c_W s_\beta \\
-M_{Z}s_W c_\beta & M_{Z}c_W c_\beta & 0 & -\mu \\ 
M_{Z}s_W s_\beta & -M_{Z}c_W s_\beta & -\mu & 0 \end{array} \right).
\end{equation}
Here, $s_W = \sin \theta_W$, $c_W = \cos \theta_W$ where $\theta_W$ is the weak mixing angle whereas 
$s_{\beta} = \sin \beta$, $c_{\beta} = \cos \beta$, where $\tan \beta = \frac{v_u}{v_d}$ refers to the ratio of the vacuum expectation value's of the up-type Higgs doublet 
($H_u$) and the down-type Higgs doublet $H_d$. 
Diagonalising the symmetric mass matrix $M^n$ using a unitary matrix $N$ lead to the neutralino mass eigenstates $\widetilde{\chi}^0_i \, ( i = 1,..,4)$,
\begin{equation}
 NM^nN^T = diag ( m_{\widetilde{\chi}^0_1},m_{\widetilde{\chi}^0_2},m_{\widetilde{\chi}^0_3},m_{\widetilde{\chi}^0_4})
\end{equation}
where $m_{\widetilde{\chi}^0_1} < m_{\widetilde{\chi}^0_2} < m_{\widetilde{\chi}^0_3} < m_{\widetilde{\chi}^0_4}$. 

The chargino mass matrix $M^c$ in the basis $(\widetilde{W}^{+},\widetilde{H}^{+}_u)$ is as follows \cite{Martin:1997ns}:
\begin{equation} \label{mc}
M^{\rm c}= \left( 
\begin{array}{cc}
M_{2} & \sqrt{2} M_W s_{\beta} \\
\sqrt{2} M_W c_{\beta} & \mu \\
\end{array} \right).
\end{equation}
Since, $M^c$ is not a symmetric matrix, we need two unitary matrices $U$ and $V$ to diagonalize the matrix. Hence,
\begin{equation}
 U^*M^cV^{-1} = diag(m_{\widetilde{\chi}^{\pm}_1},m_{\widetilde{\chi}^{\pm}_2})
\end{equation}
where $m_{\widetilde{\chi}^{\pm}_1}< m_{\widetilde{\chi}^{\pm}_2}$. In the limit where $\mu << M_1, M_2 $, 
there are two nearly degenerate Higgsino-like neutralinos along with a degenerate Higgsino-like chargino, leading to a naturally compressed spectrum. The mass eigenvalues  at the tree-level are \cite{Drees:1996pk,Giudice:1995np}
\begin{align}
\label{massHiggsino}
m_{\widetilde\chi^{\pm}_1} & = |\mu |\left(1- \frac{M_W^2 \sin2\beta}{\mu  M_2}\right) + 
\mathcal{O}(M_2^{-2})  \nonumber\\
m_{\widetilde\chi^{0}_{1,2}} & = \pm \mu - \frac{M_Z^2}{2}(1\pm \sin2\beta)\left(\frac{\sin^2
\theta_W}{M_1}+\frac{\cos^2 \theta_W}{M_2}\right)  
\end{align}
 In the presence of a light $\widetilde{G}$, the following additional 
decay modes open up for the Higgsino-like chargino and neutralinos:
\begin{align*}
  \noindent\widetilde{\chi}^{\pm}_1 \rightarrow W^{\pm} \, \widetilde{G} \,\,  \\ 
  \,\,\,\, \widetilde{\chi}^0_2 \rightarrow  h \widetilde{G}, Z \widetilde{G} \\ 
  \,\,\,\, \widetilde{\chi}^0_1 \rightarrow h \widetilde{G}, Z \widetilde{G}   
\end{align*}
 where the $Z$ boson from a neutralino decay is mostly longitudinal. 
The squared couplings of the Higgsino-like neutralinos and chargino, 
to the gravitino ($\widetilde{G}$) LSP are \cite{Djouadi:2005gj}:
\begin{align}
\begin{split}
&|g_{\widetilde{G}\widetilde{\chi}^0_iH_k}|^2 = |e_k N_{i3} +  d_k N_{i4}|^2 (M_{Pl}m_{\widetilde{G}})^{-2}, \\
&|g_{\widetilde{G}\widetilde{\chi}^{\pm}_1H^{\pm} }|^2 = (|V^2_{12}|\cos^2\beta +  |U^{2}_{12}|\sin^2\beta) (M_{Pl}m_{\widetilde{G}})^{-2}
 \end{split}
 \label{eq2}
\end{align}
where $i=1,2$ and $k = 1,2,3$ such that $H_k = h, H, A$, respectively.
The coefficients $e_k$ and $d_k$ are as below:
\begin{equation}
\begin{split}
e_1 = \cos \alpha , e_2 = -\sin \alpha, e_3 = -\sin \beta  \\
d_1 = -\sin \alpha , d_2 = -\cos \alpha, d_3 = \cos \beta \\
 \end{split}
 \label{eq3}
\end{equation} and $N_{ij}$ refer to the $(ij)^{th}$ entry of the neutralino mixing matrix $N$, $\alpha$ is the mixing angle 
between the CP-even Higgs bosons, $h$ and $H$. In the decoupling limit, i.e, $m_A >> m_h$, 
$\beta - \alpha \sim \pi/2 $ ( where, $0<\beta<\pi$ and $-\pi<\alpha<0$ ), the lightest
CP-even Higgs ($h$) boson behaves like the SM Higgs boson \cite{Martin:1997ns}.
The partial decay widths of the lightest neutralino $\widetilde{\chi}^{0}_1$ are as follows \cite{Meade:2009qv,Covi:2009bk}:
\begin{align}\label{eq:2} 
\Gamma(\widetilde{\chi}^{0}_{1} \rightarrow h\widetilde{G}) \propto |N_{14}\sin\alpha - N_{13}\cos\alpha|^2 (M_{Pl}m_{\widetilde{G}})^{-2}
\end{align}
\begin{align}\label{eq:1}
\Gamma(\widetilde{\chi}^{0}_{1} \rightarrow Z\widetilde{G}) & \propto  (|N_{11}\sin\theta_W - N_{12}\cos\theta_W|^2 \nonumber \\  + & \frac{1}{2}|N_{14}\cos\beta - N_{13}\sin\beta|^2)  (M_{Pl}m_{\widetilde{G}})^{-2}
\end{align}

The terms proportional to $N_{14}$ and $N_{13}$ denote the Goldstone couplings\footnote{See the discussion at the beginning of Sec.~\ref{LT} for the decay of the $Z$ boson and the polarization of the gauge bosons.}. In the decoupling limit,  $\sin \alpha = - \cos \beta$ and $\cos \alpha = \sin \beta$; thus Eq.~\ref{eq:2} reduces to 
\begin{equation}\label{eq:3}
\Gamma(\widetilde{\chi}^{0}_{1} \rightarrow h\widetilde{G}) \propto |N_{14}\cos\beta + N_{13}\sin\beta|^2 (M_{Pl}m_{\widetilde{G}})^{-2}
\end{equation}

The interference term between the gaugino and the Higgsino vanishes \cite{Hasenkamp:2009zz} in the above decay mode due to 
the polarization states being physical states. We note that for a Higgsino-like $\widetilde{\chi}^0_1$, $N_{13} \simeq - N_{14}$ for $\mu > 0$, whereas 
for $\mu < 0$, $N_{13} \sim N_{14}$ \cite{Meade:2009qv}. This leads to an increase in $\Gamma ( \widetilde{\chi}^{0}_1 \rightarrow h \widetilde{G})$ as evident from Eq.~\ref{eq:3}. 
 
\begin{table}[t!]
\begin{center}
 \begin{tabular}{|c|c|c|c|}
\hline
Parameters & $|\mu|$ (TeV)& sign($\mu$)& $\tan\beta$   \\
\hline
Values & 0.2-1.5&$\pm 1$ & 2-45 \\
  \hline
 \end{tabular} 
\caption{Relevant range of the input parameters for the parameter-space scan to study 
the decay probabilities of the lightest neutralino is shown. We keep 
other parameters at fixed values which include: $M_1 = 2$ TeV, $M_2 = 2$ TeV,   
$M_3 =1.917$ TeV, $M_{Q_3} = 2.8$ TeV, $M_{U_3} = 2.8$ TeV, $M_A=2.5$ TeV, $A_t = 3$ TeV and $m_{\widetilde{G}}$ = 1 keV. }
\label{tab:parameters}
\end{center}
\end{table} 

\subsection*{Branching ratios} 
\label{BR}
 \begin{figure*}[ht!]
\centering
 \includegraphics[scale=0.50]{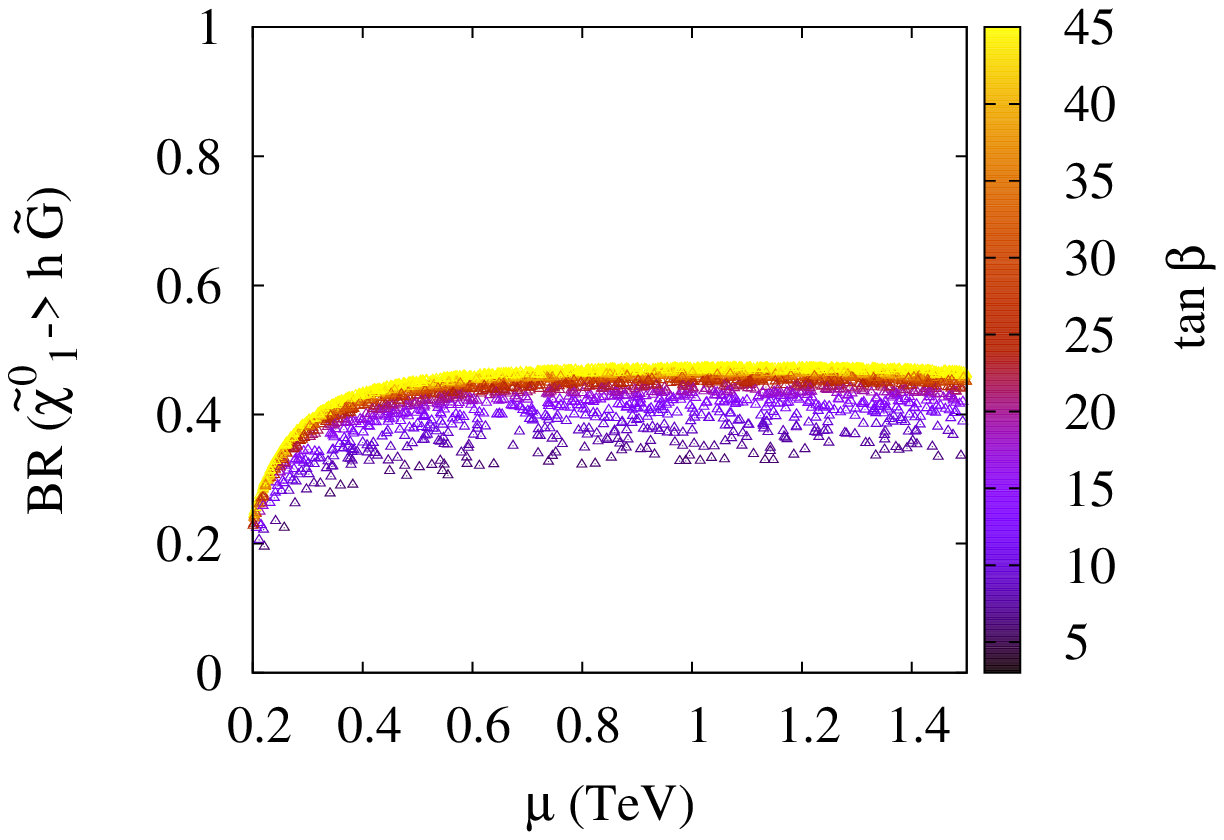} 
 \includegraphics[scale=0.50]{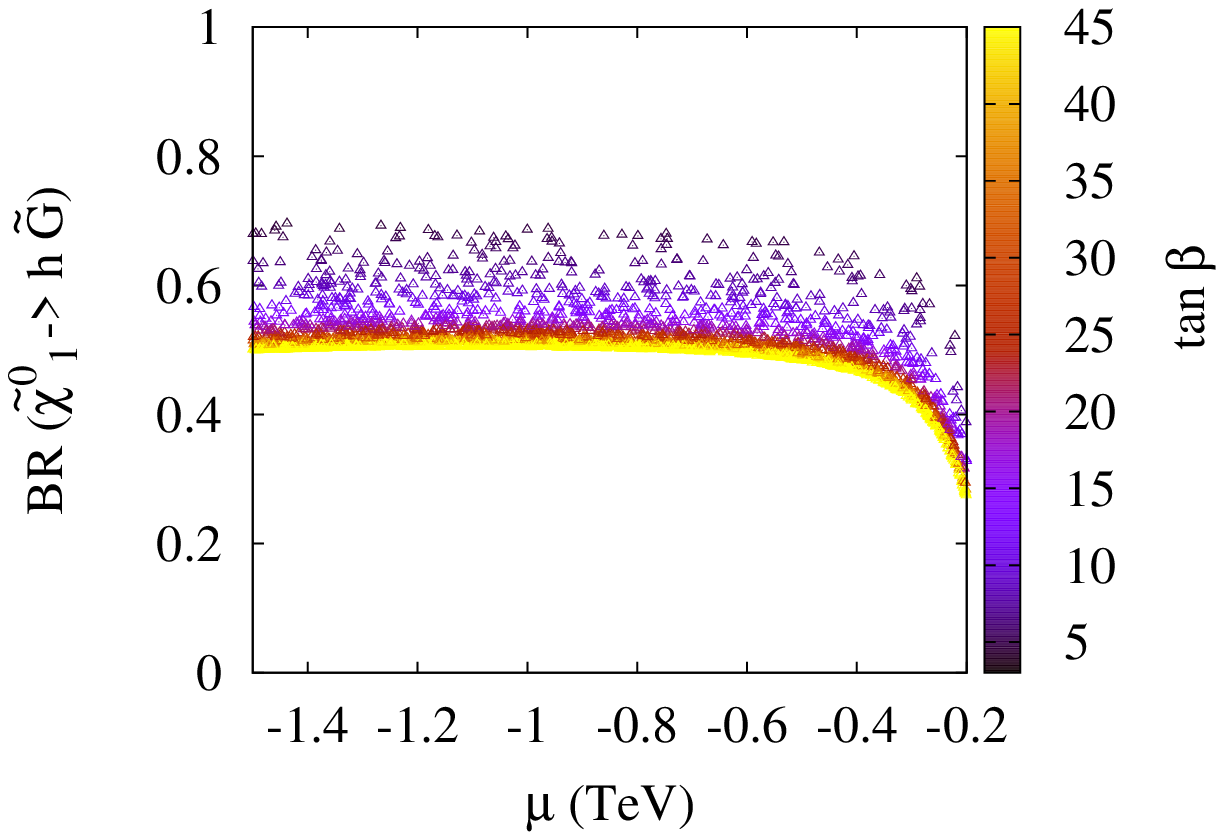}\\
 \includegraphics[scale=0.50]{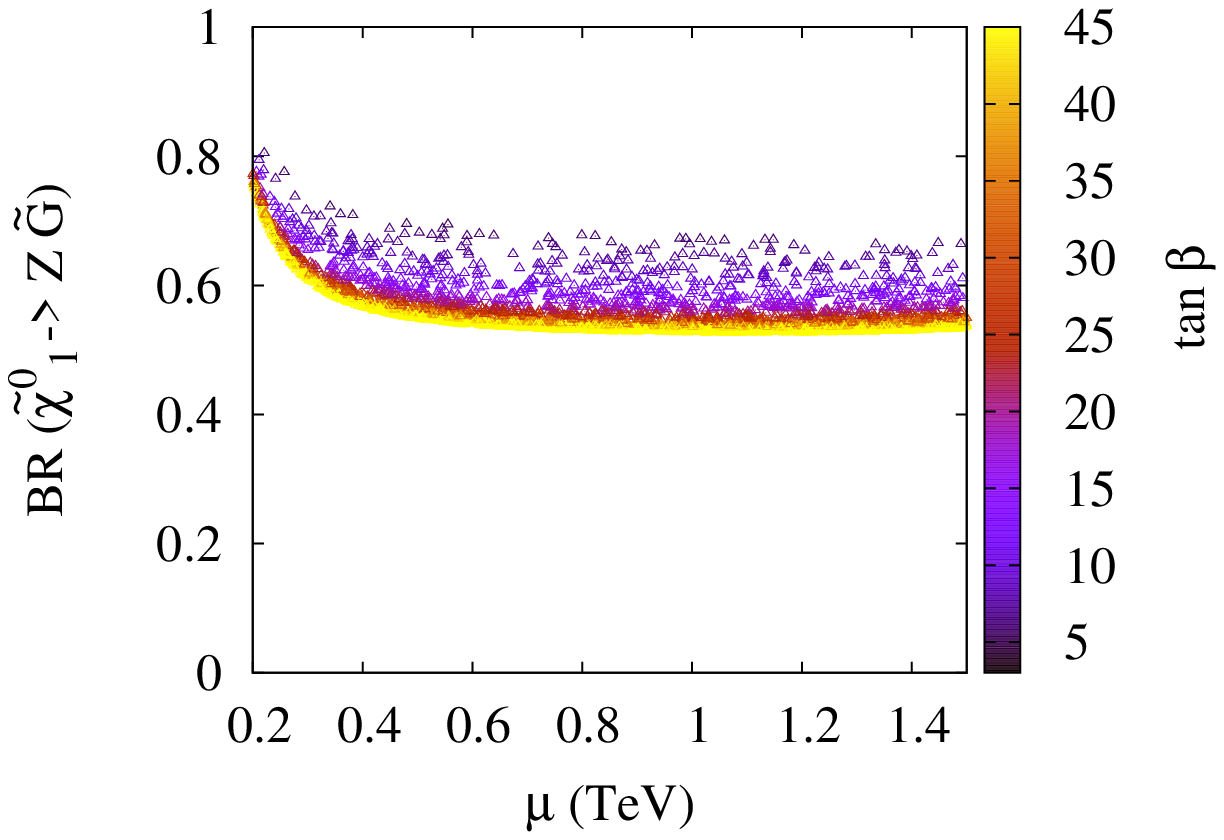} 
 \includegraphics[scale=0.50]{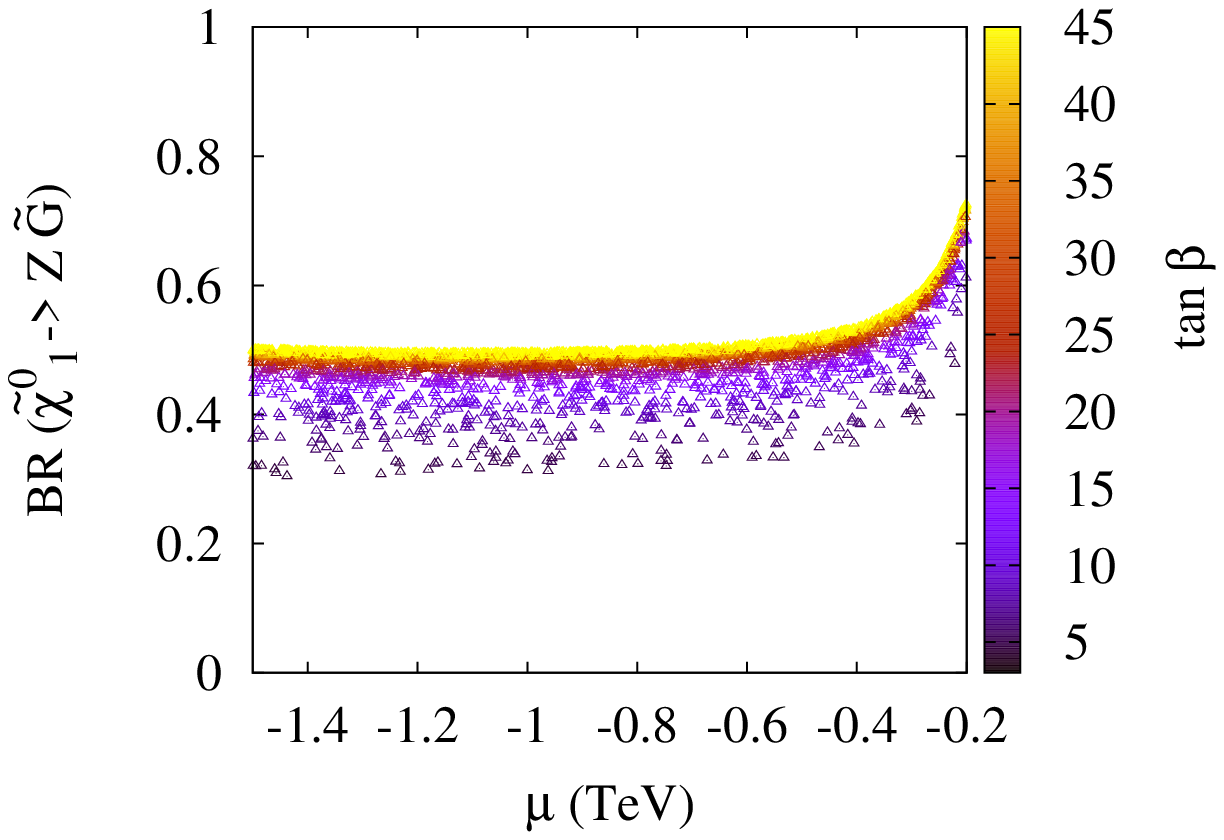}
 \caption{Variation of $\widetilde{\chi}^0_1$ decay into a Higgs (top panel) or $Z$ boson (bottom panel) along with the $\widetilde{G}$ LSP with $\mu$ and $\tan \beta$ in the colored
palette. The parameters of the scan are listed in Table~\ref{tab:parameters}.}
\label{fig:branchings}
\end{figure*} 

The decay branching ratios of the Higgsino-dominated $\widetilde{\chi}^0_1$ NLSP are governed mainly by values of $\mu$ and $\tan \beta$. Table~\ref{tab:parameters} summarizes the relevant parameter ranges for the scan performed using the package \texttt{SPheno-v3.3.6} \cite{Porod:2003um,Porod:2011nf}. In Fig. \ref{fig:branchings} we show
 the regions of the parameter space, having the different  branching ratios of the $\widetilde{\chi}^0_1$. We discuss the salient points of the parameter space as below: 
 \begin{itemize}
 
 \item The branching fractions to the Higgs and $Z$ mode decreases with an increase in the gaugino admixture in the 
NLSP at higher values of $\mu$ and $\tan \beta$ (as $\mu$ gets closer to the choice of $M_1$ and $M_2$ shown in Table \ref{tab:parameters}).  This 
defines a range of the parameter space with comparable branching ratios for the Higgs and $Z$ boson decay modes of the $\widetilde{\chi}^0_1$ NLSP.
 
 \item For $|\mu|>400$ GeV, 
 \begin{align*}
   BR(\widetilde{\chi}^0_1 \rightarrow h \widetilde{G}) \simeq BR(\widetilde{\chi}^0_1 \rightarrow Z \widetilde{G}) 
 \end{align*}
 for large values of $\tan \beta$ ($\tan \beta \geq 25$) with small differences depending on sign($\mu$). 
 For sign$(\mu) = +1$,
   \begin{align*}
 && BR(\widetilde{\chi}^0_1 \rightarrow h \widetilde{G}) \simeq 0.47, &&
 && BR(\widetilde{\chi}^0_1 \rightarrow Z \widetilde{G}) \simeq 0.53. 
  \end{align*}
whereas for sign$(\mu) = -1$,
   \begin{align*}
  &&  BR(\widetilde{\chi}^0_1 \rightarrow h \widetilde{G}) \simeq 0.54, &&
  &&  BR(\widetilde{\chi}^0_1 \rightarrow Z \widetilde{G}) \simeq 0.46. 
   \end{align*}
\item  For $\mu < 400 $ GeV and $\tan \beta \simeq 5$, 
 \begin{align*}
  && BR(\widetilde{\chi}^0_1 \rightarrow h \widetilde{G}) \simeq 0.30, &&
  && BR(\widetilde{\chi}^0_1 \rightarrow Z \widetilde{G}) \simeq 0.70 . 
 \end{align*}
due to $N_{13}=-N_{14}$ such that the $Z$ mode dominates over the Higgs mode. Whereas for $-\mu > 400$ GeV and $\tan \beta \simeq 5$,
 \begin{align*}
  && BR(\widetilde{\chi}^0_1 \rightarrow h \widetilde{G}) \simeq 0.67 ,  &&
  && BR(\widetilde{\chi}^0_1 \rightarrow Z \widetilde{G}) \simeq 0.33. 
 \end{align*}
where $N_{13}= N_{14}$  and the Higgs and $Z$ decay modes of the $\widetilde{\chi}^{0}_1$ NLSP dominate over the others respectively.
\end{itemize}

In addition, as the $\widetilde{\chi}^0_1$ becomes more gauginolike 
 the additional decay mode of $\gamma \, \widetilde{G}$ would also open up and subsequently dominate the branching 
 probabilities \cite{Matchev:1999ft}.

\section{Existing LHC limits}
 \label{constraints}
 The current bounds on the light Higgsinos as NLSP and $\widetilde{G}$ LSP are well studied at LHC for a light gravitino ($m_{\widetilde{G}}=1$ GeV 
 \cite{Aaboud:2018htj}) assuming prompt decays\footnote{Note that $m_{\widetilde{G}}=1$ GeV would correspond to a long-lived NLSP as Eq. \ref{eq1}
   suggests. As long as the mass of the gravitino would not affect the kinematics of the process, 
   it does not affect the analysis reported in
 \cite{Aaboud:2018htj}.}.
 The relevant analyses are summarized in Table~\ref{tab:limits} and we list the constraints from LHC on the Higgsinos as well as on the strong sector 
 sparticles as relevant for our study below:
   
   \begin{table} 
 \begin{center}
  \begin{tabular}{|c|c|c|c|}
   \hline
    Final state & Contributing channels& ATLAS & CMS  \\
    \hline
   $2/3/4 b + \slashed{E}_T$ & $\widetilde{\chi}^{0}_1\widetilde{\chi}^{\pm}_1 ,\widetilde{\chi}^{0}_2\widetilde{\chi}^{\pm}_1,\widetilde{\chi}^{+}_1\widetilde{\chi}^{-}_1, \widetilde{\chi}^{0}_1\widetilde{\chi}^{0}_2$&\cite{Aaboud:2018htj}& \cite{Sirunyan:2018ubx} \\
  $ \ell^+\ell^- + \slashed{E}_T$ & $\widetilde{\chi}^{0}_1\widetilde{\chi}^{\pm}_1 ,\widetilde{\chi}^{0}_2\widetilde{\chi}^{\pm}_1,\widetilde{\chi}^{+}_1\widetilde{\chi}^{-}_1,\widetilde{\chi}^{0}_1\widetilde{\chi}^{0}_2$ & &\cite{Sirunyan:2018ubx}\\
  $\geq 3 \ell +  \slashed{E}_T$ & $\widetilde{\chi}^{0}_1\widetilde{\chi}^{\pm}_1 ,\widetilde{\chi}^{0}_2\widetilde{\chi}^{\pm}_1,\widetilde{\chi}^{+}_1\widetilde{\chi}^{-}_1,\widetilde{\chi}^{0}_1\widetilde{\chi}^{0}_2$ & &\cite{Sirunyan:2018ubx} \\
   $h h + \slashed{E}_T$ & $\widetilde{g}\widetilde{g}$ & & \cite{Sirunyan:2017bsh} \\
  $4 \ell + \slashed{E}_T$& $\widetilde{\chi}^{+}_1\widetilde{\chi}^{-}_1,\widetilde{\chi}^{\pm}_1\widetilde{\chi}^{0}_2$& \cite{Aaboud:2018zeb}& \\
 $\geq2j + \slashed{E}_T$ & $\widetilde{g}\widetilde{g},\widetilde{q}\widetilde{q}$ & \cite{Aaboud:2017vwy} &  \\
  $ b\bar{b} +  \slashed{E}_T$&$\widetilde{\chi}^{0}_2\widetilde{\chi}^{\pm}_1$&\cite{Aaboud:2018ngk} & \\ 
 $1 \ell + b\bar{b} +  \slashed{E}_T$ &$\widetilde{\chi}^{0}_2\widetilde{\chi}^{\pm}_1$ &\cite{Aaboud:2018ngk} & \\ 
  $ 3 \ell  + \slashed{E}_T$ &$\widetilde{\chi}^{0}_2\widetilde{\chi}^{\pm}_1$ &\cite{Aaboud:2018ngk} & \\ 
  $  \ell^{\pm} \ell^{\pm} + \slashed{E}_T$ &$\widetilde{\chi}^{0}_2\widetilde{\chi}^{\pm}_1$ &\cite{Aaboud:2018ngk} & \\ 

  \hline
  \end{tabular}
  \caption{List of experimental searches from LHC reinterpreted for Higgsinos relevant for our current study with $\widetilde{G}$ LSP. The rate in each final state is the sum of rates in all channels listed in the corresponding rows.}
   \label{tab:limits}
 \end{center}
 \end{table}
\begin{itemize}
 \item \textbf{Higgsinos:}
 ATLAS and CMS impose stringent limits on the mass of the Higgsinos from searches involving multiple b jets/leptons 
 along with large $\slashed{E}_T$ assuming specific branching probabilities for its decay. The following are 
 the exclusion limits on the Higgsino masses \cite{Aaboud:2018htj,Sirunyan:2018ubx} :
  \begin{align*}
   BR(\widetilde{\chi}^0_1 \rightarrow h \widetilde{G}) \sim 1.0: & \,\,\, m_{\widetilde{\chi}^0_1} \leq 880 \,\, {\rm GeV \,(ATLAS)}; \\   
  & \,\,\, m_{\widetilde{\chi}^0_1} \leq 760 {\rm GeV \,\, (CMS)}. \\
   BR(\widetilde{\chi}^0_1 \rightarrow Z \widetilde{G}) \sim 1.0: & \,\,\, m_{\widetilde{\chi}^0_1} \leq 340  \,\, {\rm GeV \,\, (ATLAS)}.    
  \end{align*}
 Combined exclusion limits on the Higgsino mass from multiple searches at CMS are as follows \cite{Sirunyan:2018ubx}:
   \begin{align*}
   BR(\widetilde{\chi}^0_1 \rightarrow h \widetilde{G}) \sim 1.0: & \,\,\,  m_{\widetilde{\chi}^0_1} \leq 775 \,\, {\rm GeV \,\, (CMS)}.   \\
   BR(\widetilde{\chi}^0_1 \rightarrow Z \widetilde{G}) \sim 1.0: &\,\,\, m_{\widetilde{\chi}^0_1} \leq 650  \,\, {\rm GeV \,\, (CMS)}.    
  \end{align*}
  
 \item \textbf{Strong sector:}
  Direct limits for a massless gravitino LSP scenario are placed on strong sector sparticles with $\widetilde{G}$ LSP from 
  opposite sign dilepton + missing energy searches in ATLAS \cite{Sirunyan:2017qaj} excluding $m_{\widetilde{g}} \leq  1.8 $ TeV 
  for $m_{\widetilde{\chi}^0_1}<600$ GeV. 
  Stringent limits also arise from boosted Higgs searches \cite{PhysRevLett.120.241801} interpreted in terms of a 
  simplified scenario with a light 
  $\widetilde{\chi}^{0}_1$ LSP excluding $m_{\widetilde{g}}\leq 2.2 $ TeV for $m_{\widetilde{\chi}^0_1} = 1 $ GeV.  
  Other indirect searches which constrain the above mentioned scenario, are multijets and$/$or 
  multileptons $+$ $\slashed{E}_T$ searches \cite{Aaboud:2017vwy,Aaboud:2018zeb}, owing to the presence of $h/Z$ 
  from the NLSP decay which give rise to leptons or jets in the final state. 
\end{itemize} 

\section{Benchmarks for our analysis}
We choose representative benchmark points of the allowed parameter space to probe a low-lying Higgsino-like $\widetilde{\chi}^0_1$ NLSP with 
light $\widetilde{G}$ LSP, focusing primarily on promptly decaying $\widetilde{\chi}^{0}_1$ signals.
Our choice of benchmarks is motivated by the underlying aim of uncovering the characteristics of a 
Higgsino-like NLSP in the presence of a light $\widetilde{G}$ LSP.
Decays of the strong sector sparticles occur via the following decay modes for a keV $\widetilde{G}$:
for gluinos ($ m_{\widetilde{g}} < m_{\widetilde{q}} \leq m_{\widetilde{l}}, m_{\widetilde{\chi}^0_3}$, $m_{\widetilde{\chi}^0_4}$, $m_{\widetilde{\chi}^{\pm}_2}$), the possible decay modes to the NLSP,
\begin{align*}
& \widetilde{g} \rightarrow  t \bar{t} \widetilde{\chi}^0_1, \,\,\,\, b \bar{b} \widetilde{\chi}^0_1, \,\,\,\, t \bar{b} \widetilde{\chi}^{-}_1, \,\,\,\, 
q \bar{q}^{\prime}  \widetilde{\chi}^{\pm}_1, q \bar{q} \widetilde{\chi}^0_1. 
\end{align*}
Among these decay modes, owing to the Higgsino-like nature of the NLSP, the interaction strengths are governed by the Yukawa couplings. Hence the third generation squark channels dominate. For 
squarks ($m_{\widetilde{q}} < m_{\widetilde{g}} \leq  m_{\widetilde{l}}, m_{\widetilde{\chi}^0_3}$, $m_{\widetilde{\chi}^0_
4}$, $m_{\widetilde{\chi}^{\pm}_2}$), the possible decay modes are\footnote{The third generation squarks decay would have additional decay modes such as,
$\widetilde{t}_1 \rightarrow \widetilde{b} W^{+}$ and $\widetilde{b}_1 \rightarrow \widetilde{t} W^{-}$.}:
\begin{align*}
& \widetilde{q} \rightarrow q \widetilde{\chi}^0_1, \,\,\,\, q \widetilde{\chi}^0_2, \,\,\,\, q^{\prime} \widetilde{\chi}^{\pm}_1. 
\end{align*}
As discussed in Sec.~\ref{BR}, the dominant decay mode of the $\widetilde{\chi}^{0}_1$ NLSP is to either a Higgs or a 
$Z$ boson along with the $\widetilde{G}$ LSP which contributes to the missing energy. We wish to study the collider prospects of 
observing the final state $\geq 1 b + \ell^+\ell^- + \slashed{E}_T$ in the context of the upcoming high luminosity run of the LHC and 
explore kinematic variables reflecting the composition of the NLSP. We discuss below the characteristic features of each of 
the chosen benchmarks as shown in Table ~\ref{tab:benchmarks}.
 We also include a benchmark \textbf{BP5} similar to \textbf{BP1} with a larger branching fraction into the Higgs 
 boson and gravitino mode which would represent the low $\tan \beta$ and negative $\mu$ region of the parameter space. 

For simplicity, $M_1\simeq M_2\sim2.3-2.4$ TeV  
such that their contribution to the signal region under study (directly or via cascade decays of strong sector 
sparticles) is negligible. Among the constraints on the parameter space, light Higgs mass is within the 
range 122-128 GeV \cite{Sirunyan:2018koj,Chatrchyan:2012xdj}. In all cases, both $\widetilde{t}_1$ and$/$or $\widetilde{t}_2$ 
are heavy or the trilinear coupling $A_t$, is large to fit the lightest CP-even Higgs mass 
\cite{ARBEY2012162,Sirunyan:2018koj,Chatrchyan:2012xdj}. Also, $m_{\widetilde{\chi}^{\pm}_1}$ adheres 
to the LEP lower limit of 103.5 GeV \cite{LEPSUSYWG}.     
 \begin{table}[t]
\begin{center}
 \begin{tabular}{|c|c|c|c|c|c|c|}
 \hline
  Parameters & \textbf{BP1} & \textbf{BP2} & \textbf{BP3} &  \textbf{BP4} &  \textbf{BP5}\\
  \hline 
  $M_1$ & 2400 &800 &7000  &2300 &2400\\ 
  $M_2$ &2400  &  800  &7000  &2300 &2400\\
  $\mu$ &  800 & 2400& 700 &2250 &-800\\
  $\tan \beta$& 25  & 25 &25  &25 & 3.8\\
  $A_t$ & 3200  & 3200  &100& 3200 &3740\\
  $m_A$ & 2500 & 2500    & 2500&2500 &3000\\
  \hline
  $m_h$ & 125.3 & 125.3&127.1 &124.5 &122.2\\
  $m_{\widetilde{g}}$ & 2806.4  & 2807.1& 7271.2  &2840.1 &2663.3\\
  $m_{\widetilde{q}_L}$ & 2303.3 & 2300.2 &7156.4   &2313.3&2280.6\\
  $m_{\widetilde{q}_R}$ & 2302.2  & 2302.5 & 7155.4  &2312.5 &2283.7\\
  $m_{\widetilde{t}_1}$ & 2357.5   & 2184.8 &7057.0 &2509.1 & 1581.1\\
  $m_{\widetilde{t}_2}$ & 2340.9 & 2370.8&7104.0 &2666.0 &2271.4\\
  $m_{\widetilde{b}_1}$ & 2260.9   & 2266.4&7102.2  &2583.4 &2237.5\\
  $m_{\widetilde{b}_2}$ & 2299.0 & 2323.9& 7129.0& 2630.3 & 2295.6\\
  $m_{\widetilde{l}_L}$ & 3331.8 & 3326.8& 7337.2& 3332.6  &3329.4\\
  $m_{\widetilde{l}_R}$ & 3335.6 & 3333.7&7336.3 & 3336.3 &3334.1\\
  $m_{\widetilde{\chi}^{0}_1}$ &  810.9  & 797.9& 718.8& 2211.0  &1214.8\\
  $m_{\widetilde{\chi}^{0}_2}$ & -814.4 & 837.8 &-723.7 &-2254.8  & -1217.2\\
  $m_{\widetilde{\chi}^{\pm}_1}$ & 812.5 & 837.9&720.9 & 2223.1  & 1216.4\\
  $m_{\widetilde{\chi}^{\pm}_2}$ & 2415.7   & 2397.3&1925.9 & 2350.5 & 2420.9\\
  $m_{\widetilde{\chi}^{0}_3}$ & 2386.3   & -2394.8 &1923.6 &2290.1 &2392.2 \\
  $m_{\widetilde{\chi}^{0}_4}$ & 2415.6  &  2397.4&1925.8 & 2350.5 & 2420.9\\
  $m_{\widetilde{G}}$ (keV) &  1.0 & 1.0  &1.0& 1.0 &1.0 \\
  \hline
  $BR(\widetilde{\chi}^{0}_1 \rightarrow h \widetilde{G})$ & 0.45 & 0.0& 0.44 & 0.23 &0.27 \\
  $BR(\widetilde{\chi}^{0}_1 \rightarrow Z \widetilde{G})$ & 0.55&  0.25& 0.56 &0.75 &0.73\\
  $BR(\widetilde{\chi}^{0}_1 \rightarrow \gamma \widetilde{G})$ & 0  &0.75&0  &0.02 &0\\
  \hline
  $BR(\widetilde{\chi}^{\pm}_1 \rightarrow W \widetilde{G})$ & 0.024&  0.0& 0.003 &0.0001 &0.15\\
  $BR(\widetilde{\chi}^{\pm}_1 \rightarrow W^* \widetilde{\chi}^{0}_1)$ & 0.976&1.0&0.997 &0.9999 &0.85 \\
  \hline
 
 \end{tabular}
\caption{List of benchmarks chosen for our study. Mass parameters are in GeV 
unless specified otherwise.} 
\label{tab:benchmarks} 
 \end{center}
\end{table}
We choose the benchmarks after passing them through the public software \texttt{CheckMATE} \cite{Drees:2013wra}. 
Among the searches implemented in  \texttt{CheckMATE}, stringent constraints come from multijet searches 
by ATLAS \cite{Aaboud:2017vwy}. The benchmark points are generated using the spectrum 
generator \texttt{SPheno-v3.3.6} \cite{Porod:2003um,Porod:2011nf}.

\section{LHC Signals}
We now discuss in detail the possible LHC signals arising in the current scenario with a Higgsino-like $\widetilde{\chi}^0_1$ 
NLSP and keV $\widetilde{G}$ LSP. Strong sector sparticles pair produced at $\sqrt{s}=13$ TeV LHC cascades down to
the $\widetilde{\chi}^0_1$ NLSP along with additional jets arising from the cascade. 
In situations where the strong sector is not kinematically accessible, it is worthwhile to explore signals from the direct 
production of the low-lying Higgsinos decaying promptly to the NLSP $\widetilde{\chi}^0_1$ which then further decays to a 
Higgs/$Z$ gauge boson and the $\widetilde{G}$ LSP. As discussed in Sec.~\ref{BR}, such a scenario would lead to 
$hh/hZ/ZZ$ final states with/without extra hard jets arising from the strong sector cascade.

Motivated by the characteristics of a Higgsino NLSP spectrum, among the multifarious signatures possible we focus on a final state consisting of a Higgs and $Z$ boson along with large $\slashed{E}_T$ as the primary signature of such a scenario. In addition, to study the characteristic polarization of the $Z$ boson
coming from the decay of the NLSP we require an efficient and cleaner mode of reconstruction which can only come through the leptonic decay of the weak gauge boson.  Note that
for the hadronic decays of a Higgs and $Z$ boson contributing to the signal rates the corresponding SM hadronic background would also be significantly bigger.
We therefore choose a final state that includes at least one $b$ jet and two same-flavor opposite sign leptons along with 
$\slashed{E}_T$. Owing to the presence of leptons in the final state, this is a relatively clean channel to observe at LHC
as compared to an all hadronic final state. Since the LSP is a very light $\widetilde{G}$, the ensuing $h/Z$ from the NLSP decay
and hence, the $b$ jets and$/$or leptons have large transverse momentum ($p_T$), thereby leading to a large $\slashed{E}_T$, 
where $\vec{\slashed{E}_T} = - \vec{p}_{T_{vis}}$ (balancing the net transverse momenta, $\vec{p}_{T_{vis}}$ of the 
visible particles). No specific criteria is imposed on the number of light jets in the scenario as will be present if the signal 
arises from the decay of the squarks or gluinos to the NLSP. This is because our choice of an inclusive final state signal would 
be able to highlight the presence of a Higgsino-like NLSP irrespective of the rest of the underlying MSSM spectrum, {\it i.e.} 
with/without the strong sector placed above the low-lying Higgsinos. 
\subsection*{Signal, background and event selection criteria}
We consider the following SUSY production processes involving squarks as well as the low-lying Higgsinos to be pair produced 
when kinematically accessible: 
\begin{align*}
  p p \rightarrow \widetilde{q}_i \widetilde{q}_j, \widetilde{q}_i \widetilde{q}^{ *}_j,\widetilde{q}^*_i \widetilde{q}^*_j,\widetilde{q}\widetilde{g},
  \widetilde{\chi}^0_1\widetilde{\chi}^0_2, \widetilde{\chi}^{\pm}_1\widetilde{\chi}^0_1,\widetilde{\chi}^{\pm}_1\widetilde{\chi}^0_2,
  \widetilde{\chi}^{+}_1\widetilde{\chi}^{-}_1. 
\end{align*}
Note that the gluinos are heavier compared to the squarks and Higgsinos and their contribution to the signal is subdominant. When the signal is generated from the pair production of the strongly interacting sparticles, the final state consists of at least 
two hard jets in the $hh/hZ/ZZ$ final state along with a pair of invisible 
gravitinos which contribute to the large $\slashed{E}_T$.  Among the possible 
combinations of the decay products of $h$ and $Z$, we primarily focus on
the $\geq 1 b + \ell^+ \ell^- $ final state along with $\slashed{E}_T$. Since the $Z$ decays leptonically, 
it gives a cleaner channel and better control over the SM backgrounds as compared to a hadronic final state.  
 
  We generate the signal events in \texttt{Madgraph$\_$v5} \cite{Alwall:2014hca} using the model UFO files available from 
\texttt{Feynrules} \cite{Christensen2013}. 
Subsequently, parton level events are showered and hadronised using \texttt{Pythia} \cite{Sjostrand:2006za,Pythia8}, and a detector simulation is performed
using \texttt{Delphes} \cite{deFavereau:2013fsa}. Jets (including b   jets) are reconstructed using the anti-kT algorithm \cite{Cacciari:2008gp} using \texttt{Fastjet} \cite{Cacciari:2011ma} with minimum transverse momentum, $p_T > $ 20 GeV within a cone $\Delta R = 0.4 $. Charged leptons are reconstructed in a cone of $\Delta R=0.2$ with a maximum energy
deposit in the cone from all other particles limited to $10\%$ of the $p_T$ of the lepton. 
The significant contributions to the SM background for the given final state come from
\begin{itemize}
 \item $t \bar{t}$, $( t \rightarrow b W^+, \text{ } W^+ \rightarrow \ell^+ \nu) $
 \item $hZ +$ jets, ( $h\rightarrow b \bar{b},\text{ } Z \rightarrow \ell^+\ell^-$)
 \item  $t \bar{t} Z$,  ( $Z \rightarrow \ell^+ \ell^- )$
 \item $Z Z$, ( $Z \rightarrow b \bar{b} \text{, } Z\rightarrow \ell^+\ell^- )$
 \item $W^{\pm}W^{\mp}Z$, ($Z \rightarrow \ell^+\ell^-)$
 \item $l^+l^- b\bar{b}+\nu\bar{\nu}$ 
 \end{itemize}
Although the QCD background has a large cross section, it has a negligible 
contribution to the signal region characterized by large $\slashed{E}_T$ as well as 
effective mass ($M_{Eff}$), which helps probe the heavy mass scale of the 
SUSY particles and would serve as an effective discriminator between the SUSY signal 
and SM background. For SM background,  we perform \texttt{MLM} matching \cite{Alwall:2014hca} when needed with \texttt{QCUT}=20$-$30 GeV.

\subsection*{Primary selection criteria}
We choose the following basic selection criteria to identify leptons ($e,\mu$) and $b$ jets in the signal and background:
\begin{itemize}
 \item The charged leptons are identified with $p_T>10$ GeV and $|\eta|<2.5$.
  \item All reconstructed $b$ jets have $p_T > 30$ GeV and  $|\eta|<2.5$.
 \item Jets and leptons are isolated with $\Delta R_{ij} > 0.4$ and $\Delta R_{\ell\ell}>0.2$. 
\end{itemize}
  \subsection*{Signal analysis: At least 1$b$-jet + $\ell^+ \ell^-+\slashed{E}_T$}
We note that for the signal, since the NLSP-LSP mass gap is large, the transverse momenta carried by the 
decay products are large thereby ensuring a large amount of $\slashed{E}_T$ in the event. Fig.~\ref{fig:kinemats} shows the normalized differential distribution of a few kinematic 
variables ($M_{Eff}$ and $\slashed{E}_T$) for \textbf{BP1} and \textbf{BP4} along with the 
background. The SUSY signal distributions for the missing transverse 
energy ($\slashed{E}_T$) and effective mass ($M_{Eff}$) are widely separated from the SM 
background for \textbf{BP4} in the presence of a heavy NLSP. However, the signal 
events peak at a much lower value of $M_{Eff} \sim 850$ GeV for  \textbf{BP1}, while significant 
events of the signal are found at large $M_{Eff}$ values $\sim 2.0$ TeV for \textbf{BP4}. 
Note that for \textbf{BP4}, this is due to the high transverse momentum of the jets, and 
leptons arising from the decay cascades of the heavy $\mathcal{O}(2)$ TeV 
range sparticles. However, for \textbf{BP1} with a light NLSP, there is considerable overlap of the kinematic distributions with the background while differing in the tail of the distribution. 
This happens because the dominant contribution to the signal comes from the 
direct production of the light Higgsino sector as compared to the strong production 
cross section.  We break our analysis in two parts to study 
different scenarios that can present themselves at LHC. The signal from a heavy spectrum 
of $\mathcal{O}(2)$ TeV including the NLSP, that can only have a relevant signal contribution
through the production of strongly interacting sparticles at the LHC is optimized using cuts 
in \textit{Analysis 1} while the signal for relatively lighter electroweakino states being 
directly accessible at LHC with smaller contributions from the strong sector is analyzed in 
\textit{Analysis 2}. Appropriate cuts on the relevant kinematic variables will be crucial 
to remove SM background in the subsequent collider analyses to study the two scenarios
discussed above.

\subsection*{Analysis 1}
As a crucial part of our analysis is 
dependent on the reconstruction of the $Z$ boson in the events through the dilepton 
mode, the event rate for the signal will suffer due to the small branching fraction of the 
gauge boson to charged leptons. In addition, if we intend to reconstruct the light Higgs
boson too using double $b$-tag jets, we will end up restricting our search sensitivity significantly. We therefore need to select events using proper cuts to be able to 
identify the $Z$ boson as well as imply a Higgs like event. 
In order to select such a final state we implement the following event selection 
criterion to retain a significant amount of signal against the SM background: 

\begin{figure*}[ht!]
\centering
\includegraphics[width=2.2in,height=2.0in]{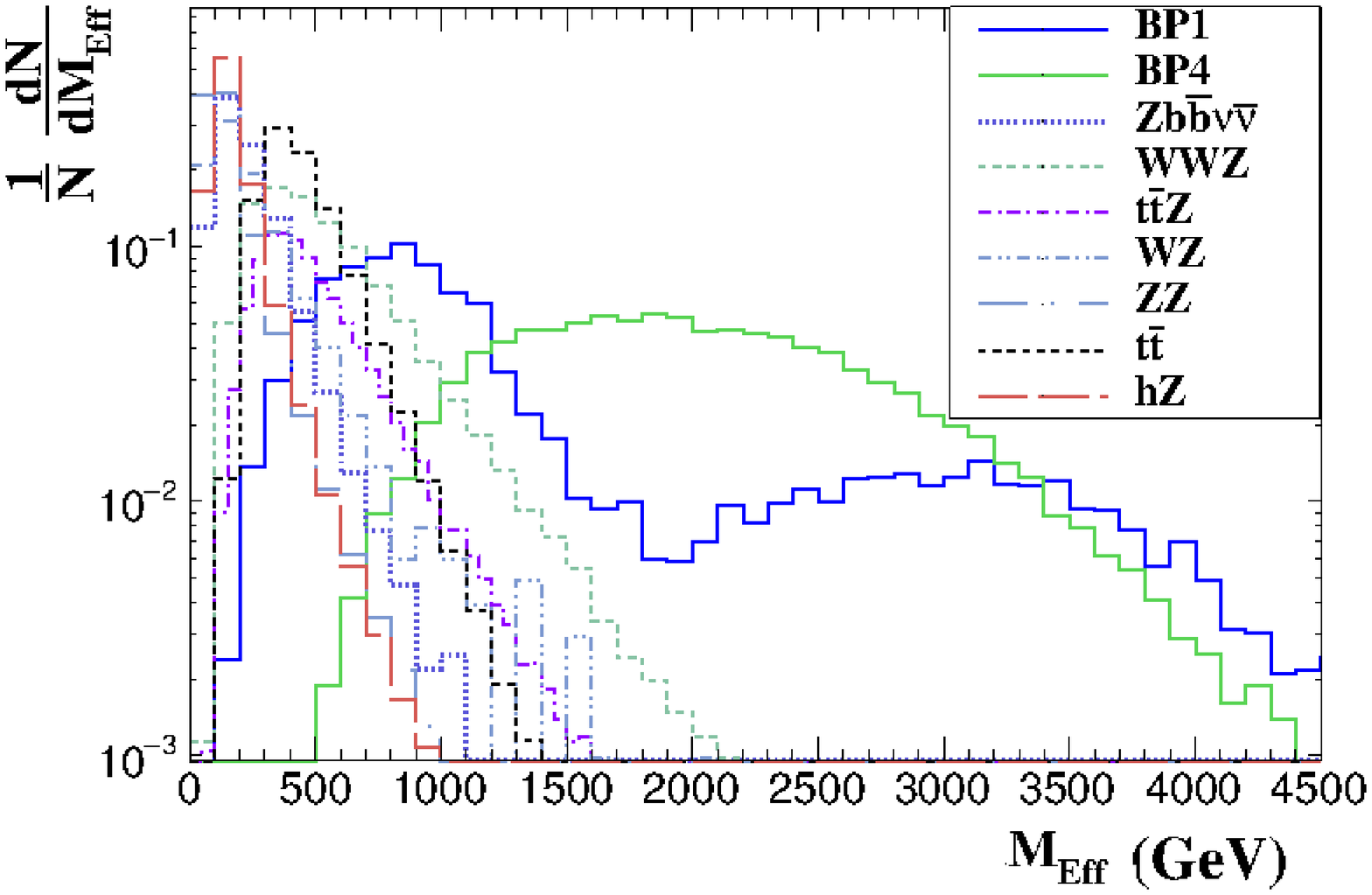}
\includegraphics[width=2.2in,height=2.0in]{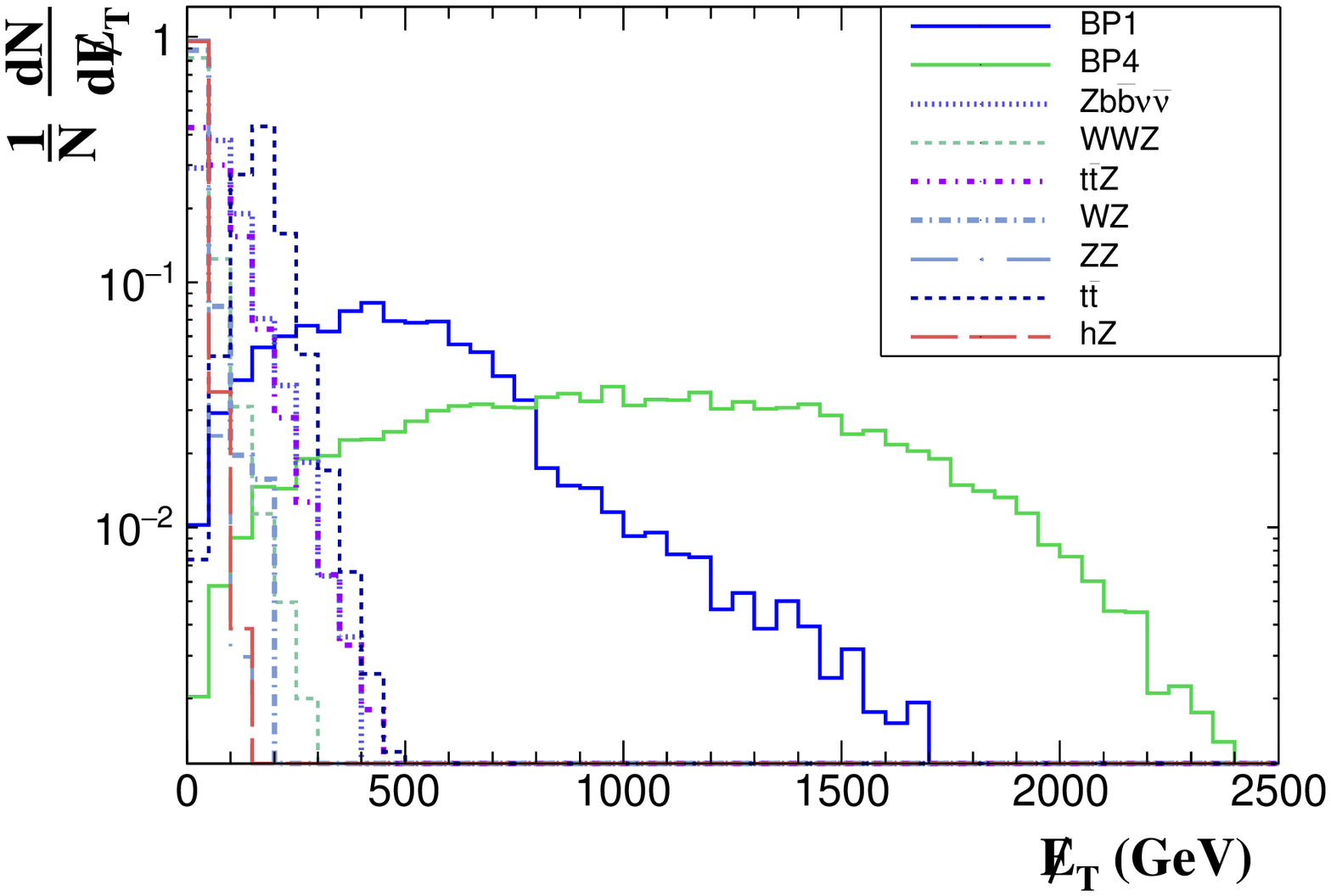}
  \caption{Distribution of few useful kinematic variables before application of any selection cuts.}
\label{fig:kinemats}
\end{figure*}

\begin{itemize}
 \item \textbf{D1}: We select a final state with two opposite sign leptons of same 
 flavor ($N_{\ell}= 2$ with $p_T > 20$ GeV) and at least one $b$ jet
 with $p_T>30$ GeV. 
   \item \textbf{D2}: To reconstruct the $Z$ boson we demand that the invariant mass 
   of a dilepton pair (opposite sign same-flavor) in the signal events is within the $Z$ 
   mass window satisfying $76<M_{\ell^+\ell^-}<106$ GeV.
 \end{itemize}
 Another kinematic variable of importance is the \textit{stransverse mass} 
 $M_{T_2}$ \cite{Cheng:2008hk}. It is reconstructed using the $p_T$ of the charged lepton 
 pair along with $\slashed{E}_T$. For SM processes such as $t \bar{t}$, 
 $M_{T_2}$ shows an end point value $\sim M_W$. For SUSY 
processes, the end point is determined by the mass difference between the NLSP and LSP. 
For a light keV scale LSP and TeV NLSP, the end point
is large compared to $M_{W}$ and can serve as an effective discriminator between 
SUSY signals and the dominant SM background subprocesses.
 \begin{itemize}
 \item \textbf{D3}: We demand a cut on the kinematic variable $M_{T_2} > 90 $ GeV to remove backgrounds from $t\bar{t}$.
\end{itemize}
The other important kinematic variable is the effective mass,  
$M_{Eff} =  p_{T} (\ell^+) + p_{T}(\ell^-)+\slashed{E}_T + \sum p_{T_i}(j)$, the scalar 
sum of the transverse momenta of the visible jets, leptons and $\slashed{E}_T$ in the event.
$M_{Eff}$ reflects the mass scale of the heavy SUSY particles and hence, is an efficient variable to suppress SM background. However the choice of a strong cut on $M_{Eff}$ would choose 
to retain contributions from a very heavy spectrum, and therefore, in this case, we focus 
on the benchmarks that contribute mainly through the production of the strong sector 
sparticles, {\it viz.}  \textbf{BP4} and \textbf{BP5}. Note that the strong sector for 
\textbf{BP1} is of similar value to \textbf{BP4} and \textbf{BP5}, and 
therefore the signal rates coming from the strong sector would be very similar. However, 
a dominant fraction of signal events would come from the light Higgsino production, and 
the cuts given below  are not particularly optimized to study \textbf{BP1}. 
The \textbf{BP3} signal, on the other hand, becomes very small. We shall discuss the  
\textbf{BP1} and \textbf{BP3} benchmarks separately in \textit{Analysis 2}. Additional cuts 
on the events are 
\begin{itemize}
 \item \textbf{D4}: Since nearly all the SUSY particles excluding the LSP are heavy for 
 \textbf{BP4} and \textbf{BP5}, a large $M_{Eff}$ is expected for the signal over the 
 SM background as shown in Fig. \ref{fig:kinemats}. We therefore demand a strong 
 cut of $M_{Eff} > 2$ TeV. This cut renders the signal for other benchmarks to a 
 relatively smaller value.
 \item \textbf{D5}: In addition, we also put a strong cut on missing transverse 
 energy, $\slashed{E}_T > 300$ GeV to further remove remaining contributions from 
 SM background processes.
  \end{itemize}
\begin{table}[h!]
\begin{center}
\small
 \begin{tabular}{|c|c|c|c|c|c|}
 \hline
  Signal & \textbf{D1} &  \textbf{D2} &  \textbf{D3} & \textbf{D4} & \textbf{D5} \\
  
  \hline
  \textbf{BP1}
   &130 &104 &91 & 39 &33 \\
  \hline 
  \textbf{BP3} &98 &83   &74    &2   &2 \\
  \hline
  \textbf{BP4} 
   &22 &17&17&16&16\\
  \hline
  \textbf{BP5} 
   & 62 & 33 & 30 & 28 & 26 \\
  \hline
   
  SM background & \textbf{D1} &  \textbf{D2} &  \textbf{D3} & \textbf{D4} & \textbf{D5} \\
  \hline
 $t \bar{t}$ &  365125  &64968   &186  & ... & ...  \\
   $h Z$& 29348& 28360&781 &1.76 &0.16  \\ 
  $ZZ$& 178581 & 172636 & 2124 & 15 & 2.3  \\
   $t \bar{t}Z$ & 3043.3&2111   &287  &6.14  &0.98   \\
   $t \bar{t}W$& 9121  &1802   & 14 & ... & ...   \\
   $WWZ$ & 159 &153  &13 &0.65  & 0.074   \\
 
  \hline
 Total background& \multicolumn{4}{|c|}{ } & 3 \\
 \hline
 
 \end{tabular}
\caption{Number of signal and background events for $\geq 1 \text{ } b + \ell^+\ell^- + \slashed{E}_T$ at $\sqrt{s}=13 $ TeV  LHC for $\mathcal{L}=$
3000 fb$^{-1}$ using cuts \textbf{D1-D5}.  Note that the total number of background events have been rounded off to the nearest integer. Cross sections for SUSY signals
have been scaled using NLO K factors  \cite{Beenakker:1996ed} and wherever available, NLO+NLL K factors \cite{Beenakker:2015rna}. Cross sections for SM background processes have been scaled using NLO K factors \cite{Alwall:2014hca} 
and wherever available, NNLO K factors \cite{LHCHXSWG1HELHCXsecs,Grazzini:2016swo,Cascioli:2014yka, 
Czakon:2013goa,LHCTop} have been used.}
\label{tab:analysis1}
\end{center}
\end{table}
We show the cut-flow result of our analysis for the signal and SM background in 
Table \ref{tab:analysis1}. As expected, the signal rates coming from a 2 TeV squark 
sector yields quite small numbers, even with an integrated luminosity of  3000 fb$^{-1}$. 
The overwhelmingly huge SM background is brought in control by primarily using the 
$M_{T_2}$ cut and is then rendered negligibly small using the combination of  
$M_{Eff}$ and $\slashed{E}_T$ cuts. We find that the sequence of cuts shown in 
Table \ref{tab:analysis1} affects the signal slightly with a suppression of the signal rate 
of less than 50\% for \textbf{BP5}. Thus, we find a significant number of SUSY 
signal events surviving the event selection. Note that while a $\sim 75\%$ suppression 
of signal events happen for \textbf{BP1}, it is still quite large compared to the SM background, unlike that for \textbf{BP3}. 

We compute the statistical significance (${\mathcal S}$) of the above signals using the formula in Eq.\ref{sig:eqn} 
and show the required integrated luminosities to observe and discover the signal in Table~\ref{tab:sig_analysis1},
\begin{equation}
{\mathcal S} = \sqrt{2 \times \left[ (s+b){\rm ln}(1+\frac{s}{b})-s\right]},
\label{sig:eqn}
\end{equation}
where $s$ and $b$ refer to the number of signal and background events respectively. 
\begin{table}[h!]
\begin{center}
 \begin{tabular}{|c|c|c|}
 \hline
 Benchmark & $\mathcal{L}$ (in $fb^{-1}$) for 3$\sigma$ excess & $\mathcal{L}$ (in $fb^{-1}$) for 5$\sigma$ excess \\
 \hline 
\textbf{BP1} &  240 (250) & 665 (695)  \\
\textbf{BP4} & 1112 (1178) & 3090 (3270)\\
\textbf{BP5} & 340 (356)& 943 (988)  \\

 \hline
 \end{tabular}
\caption{Required luminosities for observing the SUSY signal for the different benchmarks at $\sqrt{s} = 13 $ TeV LHC run. The numbers in the parentheses include 10$\%$ systematic errors in the background.}
\label{tab:sig_analysis1}
\end{center}
\end{table}
We observe that benchmarks  \textbf{BP4} and \textbf{BP5} require large integrated luminosities, whereas \textbf{BP3} with a decoupled squark 
sector is out of reach of LHC. Although \textbf{BP1} is observable at LHC, the large $M_{Eff}$ cut reduces the contribution from the light 
Higgsino sector which is directly accessible at LHC. Therefore, this analysis is more sensitive to the 
case of heavier spectra that also includes the NLSP to be quite heavy, such as \textbf{BP4} and \textbf{BP5}. 
However, with a light Higgsino sector and similar squark masses to  \textbf{BP4}, such as in  \textbf{BP1}, we are still able to get a relatively healthy number for the signal albeit after losing a large part of the signal events. 
A more optimized set of cuts is used in  \textit{Analysis 2} to study the scenario with lighter NLSP mass.  
 
Let us also comment on the prospect of multijet searches as discovery channels for our scenario. 
Using the SM backgrounds of the multijet analyses \cite{Aaboud:2017vwy} we estimate the reach of the squark masses to 
be $2.78$ TeV to achieve a 5$\sigma$ discovery at an integrated luminosity of 3000 fb$^{-1}$ at LHC.
For such a heavy spectrum the final state channel of  $\geq 1 b+ \ell^+\ell^- + \slashed{E}_T$ would not be within the LHC reach, and therefore, the multijet channel would be the best discovery channel.

\subsection*{Analysis 2}
We now focus on the signal contribution arising dominantly from the electroweak sector of sparticles  
with/without the strong sector, as in benchmark $\textbf{BP1}$ and $\textbf{BP3}$. Since 
the Higgsino sector is lighter, a strong cut on $M_{Eff}$ as used in \textbf{D4} will deplete the signal significantly in this case. Therefore, we employ a different set of cuts 
to investigate the signal region ($\geq 1$ $b +\ell^+\ell^-$ + $\slashed{E}_T$) 
arising from the low-lying Higgsino sector.
We consider the contributions from the Higgsino sector in addition to the strong sector 
 for the benchmarks in our study  when they are kinematically accessible and study the benchmarks \textbf{BP1}  and \textbf{BP3}. 
As the final state remains unchanged, the cuts implemented on both signal and background 
in {\it Analysis 1} would still be useful for background suppression. The implemented cuts therefore remain similar except for the excluded $M_{Eff}$ cut, optimized for the two benchmarks.
\begin{itemize}
 \item \textbf{E1}: As in {\it Analysis 1}, we select a final state with two opposite 
 sign leptons of same flavor ($N_{\ell}= 2$ with $p_T > 20$ GeV) and at least one 
 $b$ jet with $p_T>30$ GeV. 
 \item \textbf{E2}: To reconstruct the $Z$ boson we demand that the invariant mass 
   of the dilepton pair in the signal events is within the $Z$ 
   mass window, satisfying $76<M_{\ell^+\ell^-}<106$ GeV.
 \item  \textbf{E3}: As before, $M_{T_2}$ is an efficient cut to reduce background contributions from $t\bar{t}$ to the signal region. We demand a slightly stronger cut 
 of $M_{T_2} > 120 $ GeV in this case as it helps improve the signal-to-background 
 ratio. 
 \item  \textbf{E4}: The SUSY signal has a larger $\slashed{E}_T$ as compared to the 
 SM background. Hence, $\slashed{E}_T > 300$ GeV cut helps reduce a significant 
 part of the remnant contributions from SM background.
 
\end{itemize}
\begin{table} [h!]
\begin{center}
 \begin{tabular}{|c|c|c|c|c|}

 \hline
  \textbf{BP1} & \textbf{E1} &  \textbf{E2} &  \textbf{E3} & 
   \textbf{E4} \\
  \hline
 
  $\widetilde{\chi}^{\pm}_1\widetilde{\chi}^{\pm}_1$& 16& 13& 11&9 \\
  $\widetilde{\chi}^{\pm}_1\widetilde{\chi}^{0}_{1/2}$&65&54&47 &36 \\
    $\widetilde{\chi}^{0}_2\widetilde{\chi}^{0}_1$& 16& 14&12&9 \\
  $\widetilde{q}\widetilde{q},\widetilde{q}\widetilde{g}$ & 47&36 &28  & 26 \\
  \hline
  Total &\multicolumn{3}{|c|}{} & 80  \\
   \hline
 
   \textbf{BP3} & \textbf{E1} &  \textbf{E2} &  \textbf{E3} & %
   \textbf{E4}\\
  \hline
   $\widetilde{\chi}^{\pm}_1\widetilde{\chi}^{\pm}_1$& 33& 27&24   &18  \\
    $\widetilde{\chi}^{\pm}_1\widetilde{\chi}^{0}_{1/2}$&126&107&87  &65 \\
 
  $\widetilde{\chi}^{0}_2\widetilde{\chi}^{0}_1$& 33& 28&24  &18 \\
  \hline
 
  Total &\multicolumn{3}{|c|}{} & 101 \\

  \hline
   \textbf{SM background} & \textbf{E1} &  \textbf{E2} &  \textbf{E3} 
   & \textbf{E4} \\
  \hline
  $t \bar{t}$&365125&64968 & ... & ... \\
  $h Z$& 29348&28360&298& 0.67 \\
  $ZZ$&178581&172636&774   &6.61 \\
  $t \bar{t}Z$&3043 &2111  &151  &8.6  \\
  $t \bar{t}W$ &9121 &1802 &1   & ...  \\
  $WWZ$& 159& 153&6  &0.23  \\
  $l^+l^-b\bar{b}+\slashed{E}_T$ &2933&2905 &312  &34.7\\
  \hline
 Total & \multicolumn{3}{|c|}{ } & 51  \\
 \hline
 
 \end{tabular}
\caption{Number of signal and background events for $\geq 1 \text{ } b + \ell^+\ell^- + \slashed{E}_T$ at $\sqrt{s}=13 $ TeV  LHC for $\mathcal{L}=$
3000 fb$^{-1}$ using cuts \textbf{E1$-$E4}.  Note that  the total number of background events have been rounded off to the nearest integer. Cross sections for SUSY signals
have been scaled using NLO K factors\cite{Beenakker:1996ed} and wherever available, NLO+NLL K factors\cite{Beenakker:2015rna}. Cross sections for SM background processes have been scaled using NLO K factors \cite{Alwall:2014hca} 
and wherever available, NNLO K factors \cite{LHCHXSWG1HELHCXsecs,Grazzini:2016swo,Cascioli:2014yka, 
Czakon:2013goa,LHCTop} have been used.}
 \label{tab:analysis2}
 \end{center}
\end{table}

The cut-flow table for the signal and SM background are as shown 
in Table~\ref{tab:analysis2}.  We rely on a 
slightly stronger cut on 
$M_{T_2}$  to ensure substantial removal of the $t \bar{t}$ background while retaining the
signal events. However, other background contributions,  
such as that from $l^+l^-b\bar{b}+\slashed{E}_T$, remain. This still gives a significantly large 
event rate for the signal as compared to {\it Analysis 1}, and 
thereby allowing a $\sim(8-10)\sigma$ discovery possible with 
$\mathcal{L} = 3000$ $fb^{-1}$. Since both the benchmarks have similar branching 
fractions into the $Z$ and Higgs mode, the difference in the required integrated luminosity 
is due to the fact that the NLSP mass is heavier in \textbf{BP1} than 
in \textbf{BP3}. The required luminosity for observing a 3$\sigma$ and 5$\sigma$ significance at LHC are shown Table~\ref{tab:sig_analysis2}. 
We conclude that both \textbf{BP1} and \textbf{BP3} are well within the discovery reach of the high luminosity run of LHC. 

\begin{table}[h!]
\begin{center}
 \begin{tabular}{|c|c|c|}
 \hline
 Benchmark & $\mathcal{L}$ (in $fb^{-1}$) for 3$\sigma$ excess & $\mathcal{L}$ (in $fb^{-1}$) for 5$\sigma$ excess \\
 \hline 
 \textbf{BP1} & 310 (333) & 862 (924) \\
 \textbf{BP3} & 208  (226) & 577 (626)\\
   
 \hline
 \end{tabular}
\caption{Required luminosities for observing the SUSY signal for the different benchmarks at $\sqrt{s} = 13 $ TeV
 LHC run. The numbers in the parentheses include 10$\%$ systematic errors in the background.}
\label{tab:sig_analysis2}
\end{center}
\end{table}

We are now set to study the efficacy of the signal that we have analyzed to identify the 
nature of the NLSP and its inherent composition with respect to the gaugino-Higgsino admixture in the following section.

\section{A Distinguishing Feature: Longitudinal vs Transverse Gauge bosons}
\label{LT}

We are interested in a Higgsino-like lightest neutralino $\widetilde{\chi}^0_1$ for which 
$\widetilde{\chi}^0_1 \rightarrow h \widetilde{G}$ is an obvious decay channel. In addition, the Higgsino also couples to the imaginary parts of the neutral component of the two Higgs doublets. This leads to Higgsino-goldstone-gravitino interactions. The goldstone, on the other hand, approximates the longitudinal component of the $Z$, when the relevant energy scale is much larger than $m_Z$. Thus, for a Higgsino-like $\widetilde{\chi}^0_1$ with $m_{\widetilde{\chi}^0_1} >> m_Z$, one also expects the decay $\widetilde{\chi}^0_1 \rightarrow Z \widetilde{G}$, when the $Z$ is longitudinal. 

Although our focus is on the polarization of $Z$ boson in the context of SUSY in this work, the polarization information of vector bosons may be 
extremely useful even for non-SUSY scenarios, where a polarized gauge boson is likely to be produced from the 
decay of a heavy particle. Thus, the features of the longitudinal $Z$ boson, which will be 
discussed in detail in this work, are also applicable for other scenarios
as well. For example, the presence of longitudinal gauge bosons from heavy Higgs 
decays have been studied in earlier works in the context of the
Tevatron \cite{DUNCAN1986517}. LHC analyses have also looked at features of 
longitudinal gauge bosons in the SM \cite{Aaboud:2019gxl}.
In case an excess over SM is observed, it is of crucial importance to extend current 
search strategies to characterize BSM scenarios by studying variables sensitive to the polarization information of the gauge bosons via their decay products. 
Although there have been several studies in the context of $e^+e^-$ colliders focusing on 
studies of polarizations of the incoming electron-positron beams or polarization of the 
final state particles, there are few analogous studies with respect to the 
LHC utilizing these techniques \cite{Meade:2010ji}. The polarization of a $Z$ boson 
has been studied briefly in~\cite{Meade:2010ji} with respect to
the LHC in a similar scenario however in displaced dilepton final states arising from the 
$Z$ boson decay using the angular variable $\cos \theta^*$ discussed below.
We discuss analytically some basic variables found in the literature, which 
distinguish longitudinal and transverse gauge bosons. The differential decay rates 
for the transversely polarized and longitudinally polarized $Z$ boson in the rest frame 
of $Z$ boson are \cite{DUNCAN1986517}
\begin{equation}
 \frac{d \Gamma_{T}}{d\cos \, \theta^{*}} \propto (1 \pm \cos \, \theta^{*})^2  
 {\label{eq:transverse}}
\end{equation}
\begin{equation}
\frac{d \Gamma_{L}}{d\cos \, \theta^{*}} \propto \sin^2 \, \theta^*, 
 {\label{eq:longitudinal}}
\end{equation}
where $\Gamma_T = \Gamma(\widetilde{\chi}^{0}_1 \rightarrow Z_T \text{ } \widetilde{G})$ 
and $\Gamma_L = \Gamma(\widetilde{\chi}^{0}_1 \rightarrow Z_L  \text{ } \widetilde{G})$ 
are the partial decay widths of the $\widetilde{\chi}^{0}_1$ to 
a transverse $Z$ boson ($Z_T$) and longitudinal $Z$ ($Z_L$), boson respectively.
  \begin{figure}[t!]
 \centering{\includegraphics[width=2.7in,height=2.4in]{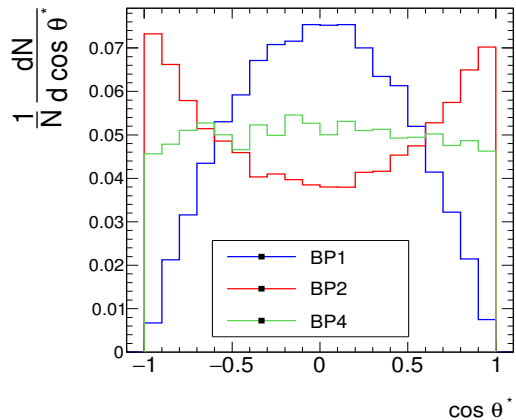}}
 \caption{Normalized distribution of $\cos \theta^{*}$ of the negatively charged lepton $(\ell^{-})$ arising from the $\widetilde{\chi}^0_1$ NLSP
 decay  at rest corresponding to the benchmarks \textbf{BP1},  \textbf{BP2}, and
 \textbf{BP4} with the isolation variable $\Delta R > 0.2 $ for the leptons.}
 \label{fig:dknlsp}
 \end{figure} 
The angle $\theta^*$ is defined as the angle the outgoing lepton (arising from the $Z$ 
boson decay) makes with the $Z$ boson in its rest frame with 
the reference direction being the boost direction of the $Z$ boson in the laboratory frame. 
The dependence of the decay width; {\it i.e.} $(1 \pm \cos \theta^{*})^2 $ corresponds 
to $ k = \mp 1$ state and $\sin^2 \theta^*$ corresponds to $k = 0$ state, 
where $k$ is the helicity of the $Z$ boson. 
To highlight the difference we choose the NLSP from a few of our benchmarks and generate a normalized distribution for $\cos \, \theta^*$ 
where the NLSP is decaying at rest and gives the $Z$ boson as its decay product. The 
simple illustration of this reconstruction is shown in Fig. \ref{fig:dknlsp} where 
{\bf BP1} represents a dominantly Higgsino-like NLSP, {\bf BP2} represents a dominantly
gauginolike NLSP while {\bf BP4} represents a comparable admixture of 
gauginos and Higgsinos in the NLSP. 

\begin{figure*}[ht!]
\centering
  \includegraphics[scale=0.28]{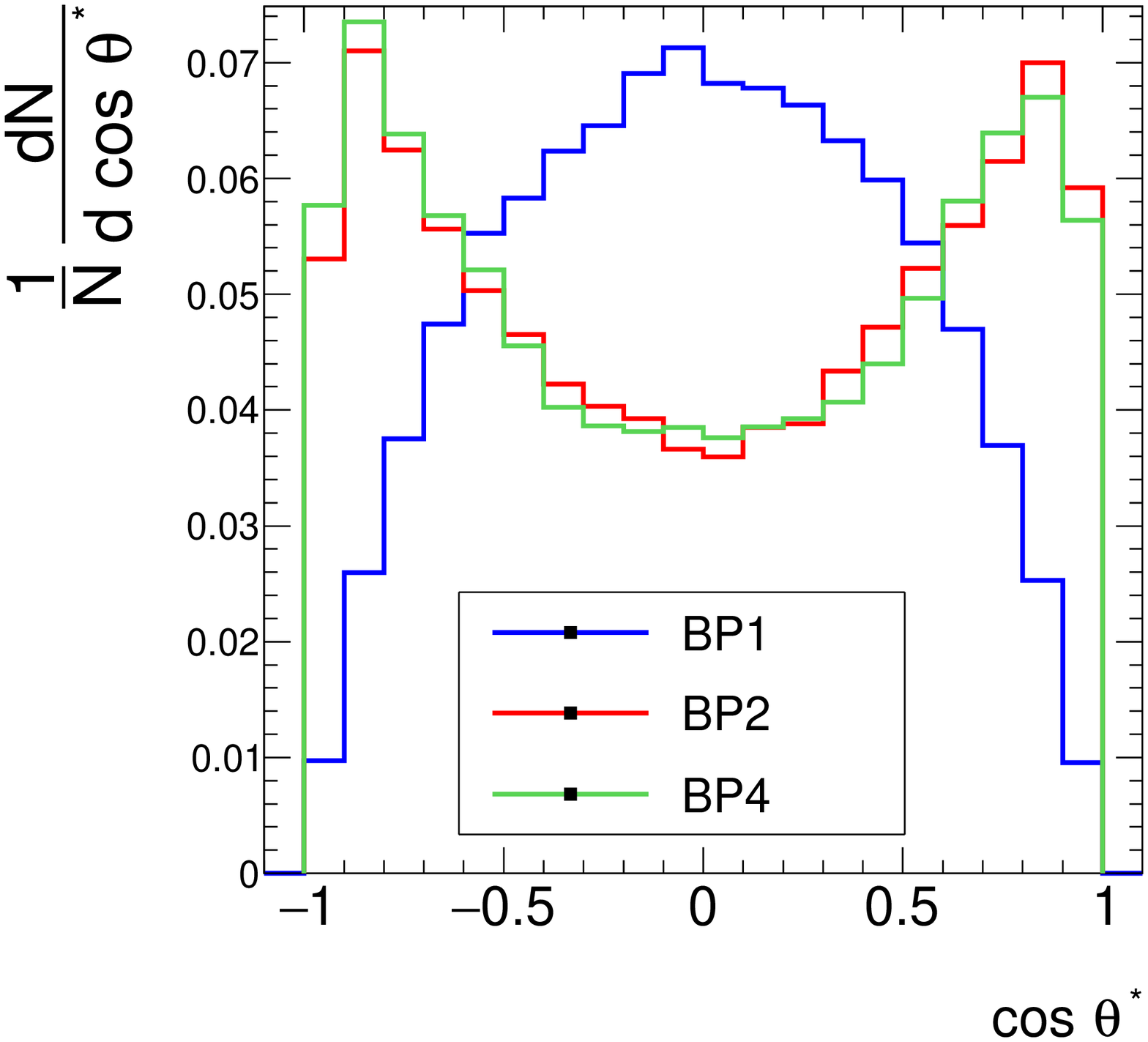} 
  \includegraphics[scale=0.23]{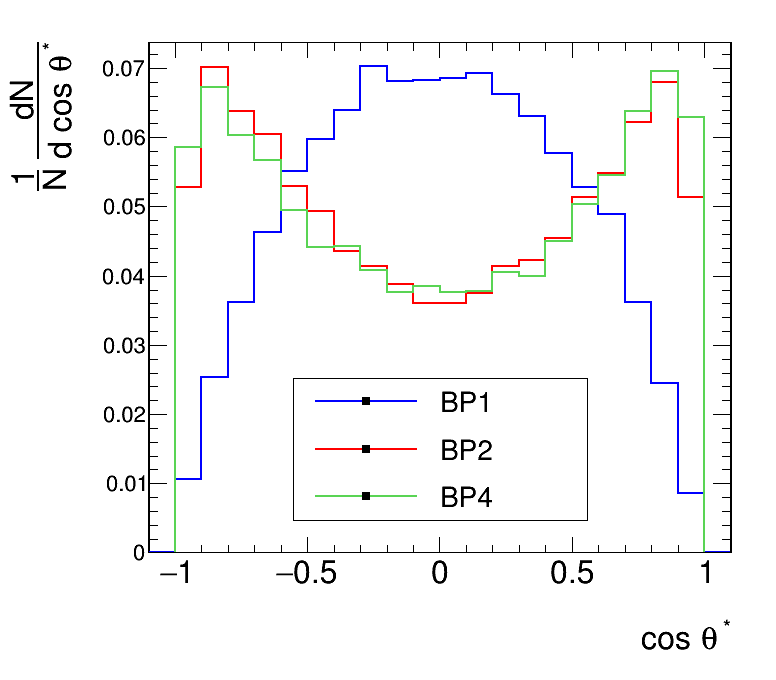} \\
  \includegraphics[scale=0.28]{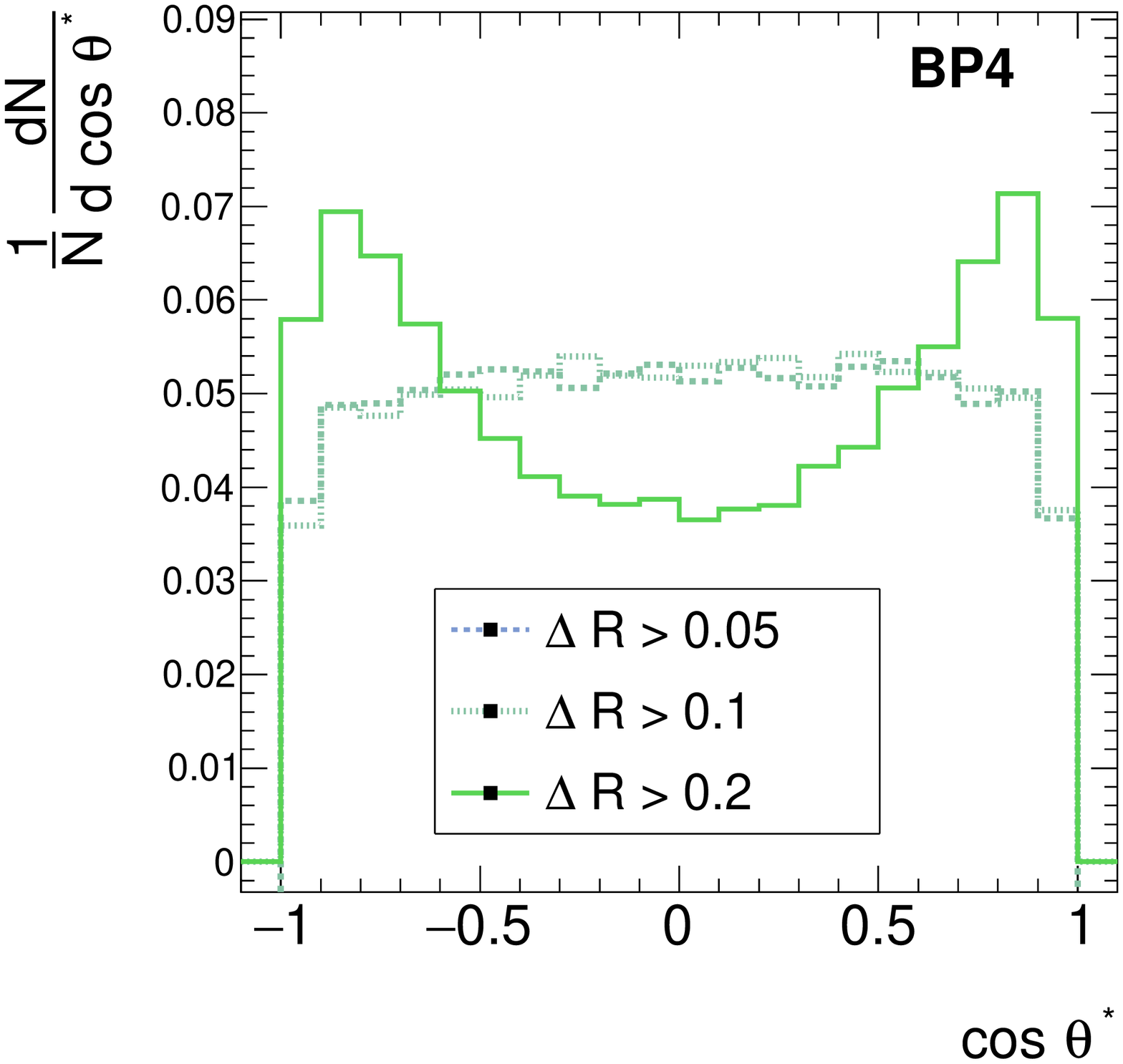} 
  \includegraphics[scale=0.25]{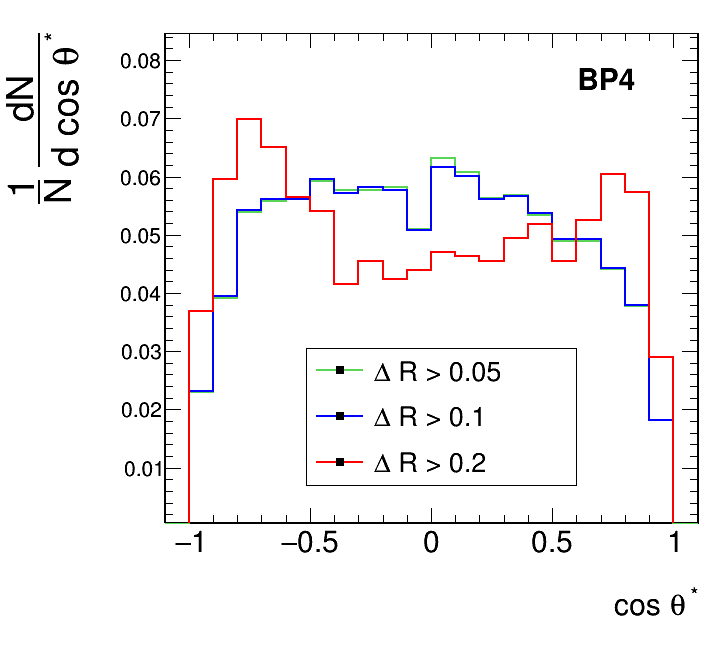}
 \caption{Normalized distribution of $\cos \theta^{*}$ of the negatively charged 
 lepton $(\ell^{-})$ arising from the $\widetilde{\chi}^0_1$ NLSP
  at the parton level (top left panel) and  after detector simulation (top right panel)  
 using \textit{Analysis 1}, corresponding to the benchmarks \textbf{BP1}, \textbf{BP2} and
 \textbf{BP4}. In the bottom panel, we present the plots for \textbf{BP4} at the parton 
 level (left) and at the detector level (right) for various $\Delta R$ values as discussed 
 in the text.}
 \label{fig:costheta}
 \end{figure*}
 We now go ahead and consider the full analysis for the signal 
 $\geq 1 \, b + \ell^+\ell^- + \slashed{E}_T$ and focus on the negatively charged lepton. 
 Note that we expect distributions in $\cos \theta^*$, as highlighted in Fig. \ref{fig:dknlsp},
 for the $Z$ polarization to be robust against the energy smearings in the detector and 
 the full detector simulation. To show this, we compare both parton-level analysis to the
 signal events obtained after detector simulations. We plot the normalized distributions 
 for $\cos \theta^*$ of the negatively charged daughter lepton of the $Z$ boson 
 in Fig.~\ref{fig:costheta} at the parton level (left) and detector level (right-panel) 
 for our benchmarks \textbf{BP1}, \textbf{BP2}  and \textbf{BP4} where one NLSP decays 
 to a $Z$ boson along with a $\widetilde{G}$.  
Recall, \textbf{BP1} has a purely Higgsino-like NLSP ($\sim 99\%$), \textbf{BP2} with purely 
gauginolike NLSP ($\sim 100\%$), and \textbf{BP4} has $ \sim 31\%$ gaugino admixture 
in the NLSP.
We observe in Fig. \ref{fig:costheta} that the distributions for the negatively 
charged lepton (for \textbf{BP1} and \textbf{BP2}) are largely similar at 
both parton and detector level simulations, 
to the expected distributions as shown in Fig. \ref{fig:dknlsp}.\footnote{Similar distributions for $\cos \theta^*$ are expected for the positively charged leptons.} For \textbf{BP4}, where the NLSP is a 
more democratic superposition of the gaugino and Higgsino states, a 
slightly flat and broad peak for $\cos \, \theta^*$ is observed. In addition, the 
NLSP mass is around 2 TeV, which results in a very boosted $Z$ boson in the final state. 
The event selection criteria can in principle have adverse effects in this case and modify the 
distributions. The most notable effect for \textbf{BP4} is 
that the distribution starts to resemble features similar to the gauginolike NLSP (\textbf{BP2}) at both parton and detector-level simulations. This we find is due to the 
fact that when the $Z$ boson is highly boosted, the pair of charged leptons coming from 
the $Z$ boson decay get more collimated with a very small opening angle. This in turn 
would mean that a larger isolation requirement for the charged leptons would lead to 
loss of events and also affect the $\cos \, \theta^*$ distribution. 

In our analysis we have used the default Delphes card using a small cone 
radius $R = 0.2$ and a maximum energy deposit in the cone being 10$\%$ of the 
$p_T$ of the lepton. A lepton-lepton isolation cut on 
$\Delta R>0.2$  reduces the peak of the $\cos \, \theta^*$ 
plot. To counter the consequent reduction in signal events,  
for \textbf{BP4} a relatively relaxed lepton identification criterion can be useful for our 
purpose. To highlight this, we identify the charged leptons with a much larger cone 
radius of $R=0.5$ for lepton identification and also demand that a large energy 
deposit with respect to the $p_T$ of the lepton is allowed in the cone ($\sim 12 \%$ for 
electrons and $25\%$ for muons).  The distribution still  retains the gauginolike behaviour 
for an isolation of $\Delta R>0.2$ as in the parton level but starts 
agreeing with the Higgsino-like feature (as in the parton-level case) when the 
separation between the charged leptons is chosen to be loose with $\Delta R>0.05$ or 
$\Delta R>0.1$ as can be seen in the bottom-right panel of Figure \ref{fig:costheta}. 

The qualitative differences observed in the distributions of the negatively charged lepton as the gaugino admixture 
increases in the NLSP amongst the three cases may be effectively captured by defining asymmetry variables in 
$\cos \theta^*$ which could clearly discriminate between a  longitudinal and transverse $Z$ boson.
Taking a cue from the features of $\cos \theta^*$, we construct a variable which highlights this difference through 
an asymmetry amongst the observed $\cos \theta^*$ values for the Higgsino-like and gauginolike NLSP. 
The asymmetry variable, $C_{\theta_Z}$, as defined in Eq.~\ref{eq:ctheta}, serves to enhance the features of the
longitudinally polarized $Z$ in comparison to the transversely polarized $Z$ such that they would be less affected if detector simulation effects smear the polarization dependence of 
the angular or energy observables. We define
\begin{align} \label{eq:ctheta}
 C_{{\theta_Z}} = \frac{N_A- N_B-N_C }{N_A+N_B+N_C}
\end{align}
where $N_I$'s stand for events whereas the subscript $I=A,B,C$ represent the angular regions in $\theta^*$ given by 
$A = [\pi/3,2\pi/3] $, $ B=[0,\pi/3]$, and $C = [2\pi/3,\pi]$.
The numerator focuses only on the asymmetry features while the denominator is the 
total number of events for $-1< \cos \theta^*<1$.
Based on the construction of $C_{\theta_Z}$, a positive value is indicative of a 
Higgsino-like NLSP whereas negative values indicate a gauginolike NLSP. 
Since $C_{\theta_Z}$ is the normalized difference in the
number of events corresponding to $|\cos \theta^*| < 0.5$ and $|\cos \theta^*| > 0.5$, 
a Higgsino-like NLSP gives larger events around 
$\cos \theta^* = 0$ as $N_A > (N_B+N_C)$, whereas for the gauginolike NLSP the distribution peaks around $\cos \theta^* \sim \pm 0.8$ {\it i.e.} $(N_B+N_C)>N_A$. Therefore, the latter shows a negative sign as compared to the former. We  list the values of $C_{\theta_Z}$ for cases when the NLSP decays at rest ($C_{\theta_Z}^{\rm \, rest}$) and 
compare this with parton-level ($C_{\theta_{Z}}^{\rm \, parton}$) results and full detector-level simulation ($C_{\theta_Z}$)
of our {\it Analysis 1}  in Table~\ref{tab:CZ}, for the benchmarks \textbf{BP1} and \textbf{BP2} (which correspond to an almost pure Higgsino and pure gaugino compositions respectively).\footnote{The results in Table~\ref{tab:CZ} are produced by using only 
squark pair 
production. However, the generic feature remains unchanged even when all production modes are included.} 
\begin{table*}[ht]
 \begin{center}
  \begin{tabular}{|c|c|c|c|c|c|c|c|}
  \hline
  Benchmark & $m_{\widetilde{\chi}^{0}_1}$ & Higgsino  & Gaugino & $C_{\theta_Z}^{\rm \, rest}$ & $C_{\theta_{Z}}^{\rm \, parton}$ & $C_{\theta_{Z}}$ & $C_{Z}$   \\
  &  (GeV) & admixture & admixture & & & & \\
  & & ($\%$) & ($\%$) & & & & \\ 

  \hline
   \textbf{BP1}& 810.9 &99.83 & 0.17 &0.38& 0.33  & 0.35 $\pm$ 0.16 &  0.38$\pm$ 0.16 \\
   \textbf{BP2} & 797.9 &0.05& 99.95& -0.18& -0.08& -0.05$\pm$ 0.21 &-0.02 $\pm$ 0.21\\
   \hline  
   \end{tabular}
\caption{Variation of the asymmetry variables $C_{\theta_Z}^{\rm \, rest}$, $C_{\theta_{Z}}^{\rm \, parton}$, $C_{\theta_Z}$ and $C_{Z}$
as defined in the text, at the parton level and detector level after cuts \textbf{D1-D5} for benchmarks \textbf{BP1} and \textbf{BP2}.
The numbers for $C_{\theta_Z}$ and $C_{Z}$ include both signal and background in the computation of the observables and its associated 1$\sigma$ error 
for $\mathcal{L}=3000 fb^{-1}$.}  
 \label{tab:CZ}
 \end{center}
 \end{table*}

Note that the values of $C_{\theta_Z}$  are in good agreement with the parton level $C_{\theta_{Z}}^{\rm \, parton}$ results.  
For the pure Higgsino-like NLSP (\textbf{BP1}), $C_{\theta_Z}$ is large and positive, 
whereas the pure gauginolike NLSP (\textbf{BP2}) shows a negative value.
We find that the $C_{\theta_Z}$ value starts decreasing as the gaugino admixture in the NLSP is increased when compared to \textbf{BP1}.  
Thus, with increasing gaugino admixture the asymmetry value becomes negative as shown for \textbf{BP2}. 
The most notable change is observed for the \textbf{BP4} with an intermediate gaugino-Higgsino admixture.  $C_{\theta_Z}$ value is $\sim 0.021$ when the NLSP decays at 
rest with the small positive value still hinting at a larger Higgsino admixture.  
However, it  turns negative for the analysis where the NLSP appears from cascade decays 
of the squark, both at the parton and the detector level owing largely to the effect of 
isolation cuts and detector smearing effects, which modify the $\cos \theta^*$ distribution 
as seen in Fig~\ref{fig:costheta} and discussed earlier. We note that the 
$C_{\theta_{Z}}^{\rm \, parton}$ value becomes positive giving $C_{\theta_{Z}}^{\rm \, parton}=0.04, \, 0.05$ for the loose 
isolation requirement and identification of the charged lepton with $\Delta R > 0.05, \, 0.1$ 
as against $C_{\theta_{Z}}^{\rm \, parton} =-0.214$ for the tighter isolation cut of 
$\Delta R >0.2$. We expect that the same would be true when the events are passed through detector simulations, which would be consistent with observations made in the 
lower panels of Fig. \ref{fig:costheta}.  

An additional kinematic feature that can be used to study the polarization of the $Z$ boson which in effect highlights the
composition of the NLSP is the charged lepton energy. Among others, the ratio of the energy carried by the charged lepton
and antilepton also show a dependence on the polarization of the $Z$ boson with an energy $E$, 
via dependence on the angle $\theta^*$. The energy ($E_\ell$) of the leptons in the laboratory frame \cite{DUNCAN1986517} follows: 
\begin{equation}
 E_{\ell}\propto \frac{E}{2}(1 \pm \beta \cos \, \theta^*)
\end{equation} 
 Using this we define two kinematic observables $Z_D$ and $Z_R$ (variations of such  observables  have been pointed out in earlier papers  \cite{Brooijmans:2018xbu,Bendavid:2018nar} using jet substructure 
to study hadronic final states),
\begin{align}
  Z_D = \frac{|E_{\ell^-}-E_{\ell^+}|}{E_{\ell^-}+E_{\ell^+}} ; &&
  Z_R = \frac{E_{\ell^-}} {E_{\ell^-}+E_{\ell^+}}  \,\,\, .
   \label{eq:Zij}
\end{align}
\begin{figure*}[ht!]
  \centering
 \includegraphics[scale=0.31]{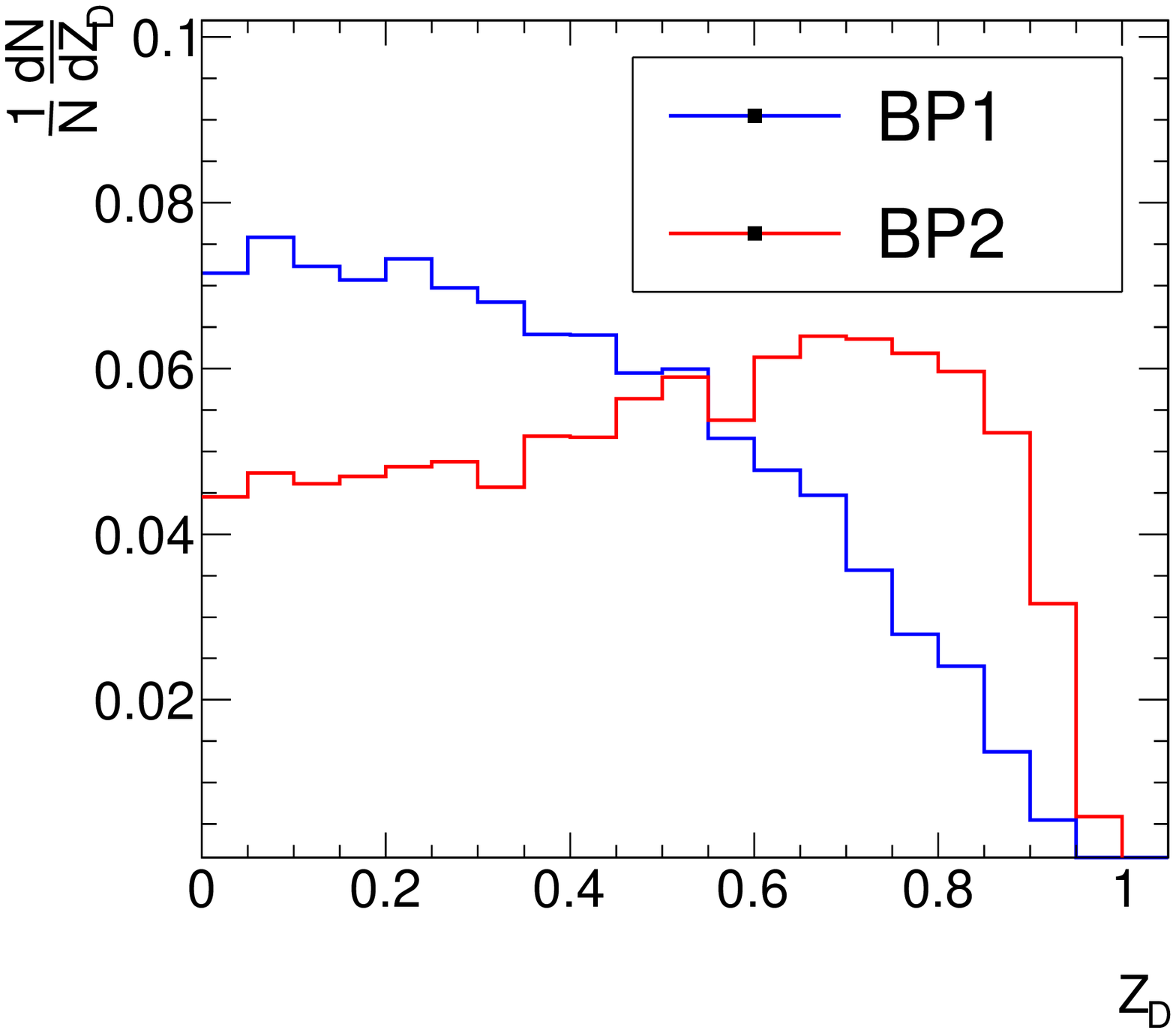} 
  \includegraphics[scale=0.35]{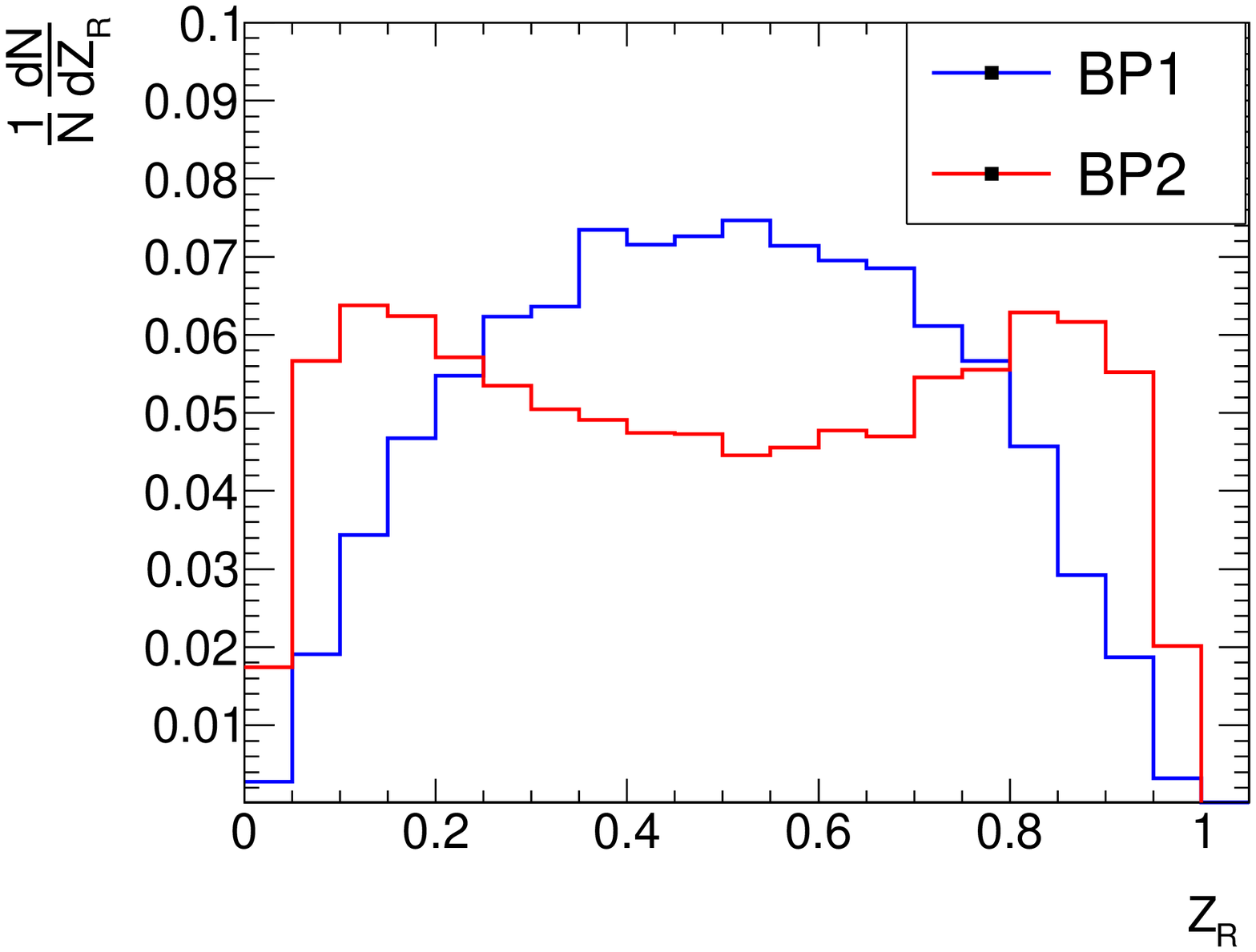} 
  \caption{Normalized distributions of the kinematic variables $Z_D$ and $Z_R$ as defined in the text for distinguishing 
  between a Higgsino and gauginolike $\widetilde{\chi}^{0}_1$ NLSP  before cuts \textbf{D1$-$D5} The variables are as 
  defined in the text. Here, we have plotted the observables for the process $\widetilde{q}\widetilde{q}$ with one of 
  the squarks decaying as:
 $\widetilde{q} \rightarrow q \widetilde{\chi}^{0}_1 \rightarrow q Z \widetilde{G}, Z \rightarrow \ell^+ \ell^-$.}
 \label{fig:energyDist}
\end{figure*}

We study the usefulness of these observables using simple cuts on kinematic variables 
after detector simulation effects are taken into account. For a predominantly longitudinal $Z$ boson, there is an equal 
sharing of energy of the parent among its daughter particles whereas for a transverse $Z$ boson, the energy sharing 
is unequal. The asymmetry is evident in Fig~\ref{fig:energyDist} where the Higgsino-like NLSP peaks at $Z_D=0.1$ as 
compared to $Z_D=0.8$ for the gauginolike case. 
Similar effects are observed in the variable $Z_R$, which denotes the fraction of net leptonic energy carried away by 
the negatively charged lepton.
The ratio  peaks at $Z_R \simeq 0.5$ for \textbf{BP1} as compared to $Z_R \simeq 0.1$ and $Z_R \simeq 0.8$ for 
the \textbf{BP2}, since for the former case, the leptons mostly have equal energy sharing, whereas unequal energy 
sharing occurs for the latter case. We define an asymmetry variable similar to $C_{\theta_{Z}}$, 
now referred to as $C_{Z}$ to capture the asymmetry in the values of $Z_D$ at the detector level,
\begin{equation} 
C_Z = \frac{N_A-N_B}{N_A+N_B}
 \end{equation}
where $N_A$ refers to the number of events for $Z_D<0.5$ and $N_B$ represents events for $Z_D> 0.5$ respectively.
We list the $C_Z$ values in Table~\ref{tab:CZ} and observe that $C_{Z}$ is positive for the Higgsino-like NLSP 
and negative for gauginolike NLSP. Note that the effect observed for the highly boosted $Z$ boson in $C_{\theta_{Z}}$ 
also shows up for $C_Z$ highlighting the consistency and importance of the isolation of the charged leptons.
The statistical uncertainty in the observed asymmetry has been shown in 
Table~\ref{tab:CZ} and is calculated using \cite{James:2006zz}
\begin{equation}
 \sigma(C) = \frac{\sqrt{1-C^2}}{\sqrt{N}}
\end{equation}
where $C = C_{\theta_Z}, C_Z$, while $N$ represents the total number of events. 
The uncertainty goes down as $ 1/\sqrt{N}$, which means that a larger luminosity would help in improving the statistical significance; 
therefore, we emphasize that the distribution of $\cos \theta^*$ 
arising from the 
$Z$ boson decay as well as the associated asymmetry variables, $C_{\theta_{Z}}$ and $C_{Z}$ prove quite useful in 
identifying the nature of the NLSP. The distinctive features of the variables discussed for distinguishing a longitudinal 
and transversely polarized $Z$ boson are also applicable for new physics scenarios where a polarized gauge boson 
is likely to be produced, and therefore, can prove very important in studying BSM physics.

\section{Summary and Conclusions}

 In this work, we have considered Higgsino-like NLSP in the presence of a light keV $\widetilde{G}$ LSP
 in the framework of phenomenological MSSM. The keV scale $\widetilde{G}$
 serves as a warm dark matter candidate, significantly relaxing constraints from dark matter searches
 on the MSSM spectrum and thereby allowing low $\mu$ parameter values. In addition, presence of a light $\widetilde{G}$ allows decay of 
  the NLSP
 to a Higgs/$Z$ boson and $\widetilde{G}$ leading to hard $b$ jets and charged leptons in the final state
 along with large $\slashed{E}_T$ carried away by the $\widetilde{G}$. Such a scenario has been extensively explored by experiments,
 including the LHC with a primary focus on the low-lying electroweak sector, leading to
 stringent constraints on the parameter space. The question that one ventures to answer in this study is as follows:
 \textit{What are the future prospects of detecting a Higgsino-like $\widetilde{\chi}^0_1$ NLSP at LHC? If detected, how can we ascertain the nature of the NLSP?} 
 
 We address this question by studying a specific final state: $\geq 1 
 b$ $+$ $ \ell^+\ell^-+ \slashed{E}_T$ at $\sqrt{s}=13 $ TeV 
 motivated by the presence of at least one $b$ jet from the Higgs boson and an opposite sign same-flavor lepton pair 
 from the $Z$ boson decay besides large $\slashed{E}_T$. 
 We choose a few representative benchmark points encompassing a light and heavy Higgsino sector with/without strong 
 sector sparticles within the reach of LHC. We find that such a signal is discoverable in the upcoming runs of the high 
 luminosity LHC after suitable cuts are applied. It is important to emphasize that such a semileptonic channel will prove 
 crucial in identifying  the nature of the NLSP, being relatively clean compared to an all hadronic final state which may have
 a better discovery prospect. Thus simultaneous use of both channels could be advocated for the purpose of discovery and
 identifying the nature of the NLSP. We focus on the presence of a dominantly longitudinal $Z$ boson arising from the 
 decay of a Higgsino-like NLSP owing to the presence of the Goldstone boson as the longitudinal mode 
 of $Z$ after electroweak symmetry breaking.  This is quite a striking identification criteria if observable, for a Higgsino-like 
 NLSP in sharp contrast to a dominantly gauginolike NLSP, which would dominantly decay to a transversely polarized $Z$. 
 It is thus  important to characterize the features of the longitudinally polarized $Z$ boson to ascertain the composition 
 of the parent NLSP. The effects of polarization of the $Z$ boson are carried by its decay products,
namely, the leptons through their angular distributions. We construct several kinematic variables using the 
negatively charged lepton as a reference and highlight its importance in observing the polarization of the parent gauge boson. 
We also propose new variables which utilize the observed asymmetries between the angular variables for the charged  
lepton coming from a parent longitudinal and transverse $Z$ boson. We do a full detector level simulation of the events and study the asymmetries that show the characteristic features of a longitudinal $Z$ boson and observe substantial differences 
between a Higgsino and gauginolike NLSP. This highlights the robustness of the constructed asymmetries.
Our analysis is applicable to other BSM scenarios which predict preferential production of longitudinally polarized gauge bosons
as a consequence of the equivalence theorem.

\begin{acknowledgments}
The work is partially supported by funding available 
from the Department of Atomic Energy, Government of India, for the Regional 
Centre for Accelerator-based Particle Physics (RECAPP), Harish-Chandra Research Institute(HRI). The research of JD is also partially supported 
by the INFOSYS scholarship for senior students at the Harish-Chandra Research Institute. We thank A. Ghosh, P.Konar, S. Mondal, and T. Samui for helpful comments, and K. Hagiwara for illuminating discussions.
\end{acknowledgments}

 \bibliographystyle{unsrt}
 \bibliography{HiggsinoGrav}
  
\end{document}